\documentclass[journal]{IEEEtran}
\usepackage{amsmath,amsfonts}
\usepackage{algorithmic}
\usepackage{algorithm}
\usepackage{array}
\usepackage{textcomp}
\usepackage{stfloats}
\usepackage{url}
\usepackage{verbatim}
\usepackage{graphicx}
\usepackage{cite}
\usepackage{multirow}
\usepackage{makecell}
\usepackage{subcaption}
\usepackage{booktabs}
\usepackage{colortbl}
\usepackage{hyperref}
\usepackage{tablefootnote}
\usepackage{bm}

\usepackage[table,xcdraw,dvipsnames]{xcolor}
\begin{document}

\title{A Large-Scale Evaluation of Speech Foundation Models}

\author{
Shu-wen~Yang,
Heng-Jui~Chang\thanks{$^*$Equal contribution; sorted alphabetically}$^{*}$,
Zili~Huang$^{*}$,
Andy~T.~Liu$^{*}$,
Cheng-I~Lai$^{*}$,
Haibin~Wu$^{*}$,
Jiatong~Shi,
Xuankai~Chang,
Hsiang-Sheng~Tsai,
Wen-Chin~Huang,
Tzu-hsun~Feng,
Po-Han~Chi,
Yist~Y.~Lin,
Yung-Sung~Chuang,
Tzu-Hsien~Huang,
Wei-Cheng~Tseng,
Kushal~Lakhotia,
Shang-Wen~Li,
Abdelrahman~Mohamed,
Shinji~Watanabe,
Hung-yi~Lee
}



\maketitle


\begin{abstract}

The foundation model paradigm leverages a shared foundation model to achieve state-of-the-art~(SOTA) performance for various tasks, requiring minimal downstream-specific modeling and data annotation. This approach has proven crucial in the field of Natural Language Processing (NLP). However, the speech processing community lacks a similar setup to explore the paradigm systematically. 
In this work, we establish the Speech processing Universal PERformance Benchmark (SUPERB) to study the effectiveness of the paradigm for speech.
We propose a unified multi-tasking framework to address speech processing tasks in SUPERB using a frozen foundation model followed by task-specialized, lightweight prediction heads.
Combining our results with community submissions, we verify that the foundation model paradigm is promising for speech, and our multi-tasking framework is simple yet effective, as the best-performing foundation model shows competitive generalizability across most SUPERB tasks.
For reproducibility and extensibility, we have developed a long-term maintained platform that enables deterministic benchmarking, allows for result sharing via an online leaderboard, and promotes collaboration through a community-driven benchmark database to support new development cycles.
Finally, we conduct a series of analyses to offer an in-depth understanding of SUPERB and speech foundation models, including information flows across tasks inside the models, the correctness of the weighted-sum benchmarking protocol and the statistical significance and robustness of the benchmark.

\end{abstract}


\section{Introduction}
\label{section:introduction}

\begin{figure}[!t]
\centering
\includegraphics[width=3.3in]{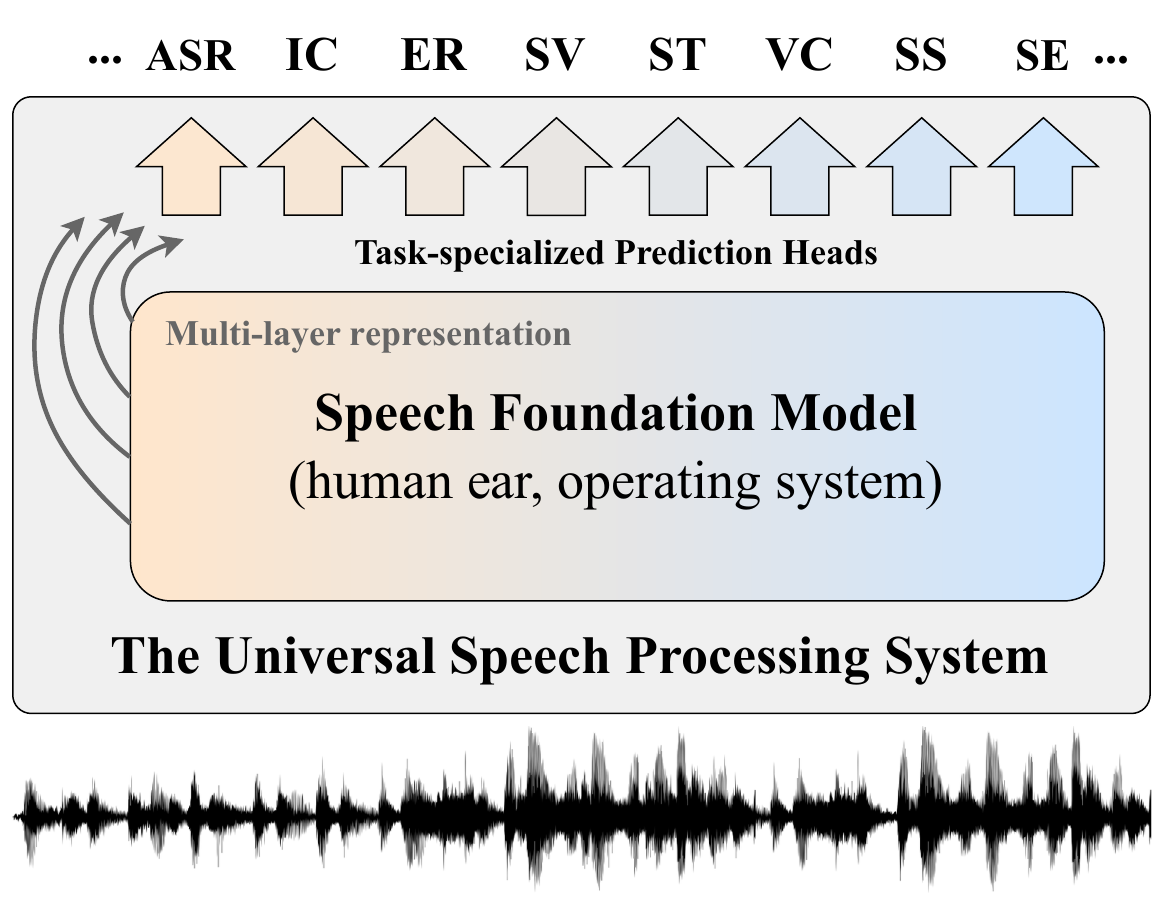}
\caption{
The diagram depicts a speech foundation model vital for a universal speech processing system. It processes waveforms from diverse real-world applications into high-level representations for predicting various speech tasks. Refer to the text for details on the abbreviations related to different speech tasks.
}
\label{fig:speech_foundation_model}
\end{figure}

\IEEEPARstart{D}{eveloping} well-performing deep learning networks has become costly, involving data collection, modeling, computing power, and training time.
The repetition for each specific use case is both time-consuming and cost-prohibitive for researchers and has a serious environmental impact~\cite{parcollet21_interspeech}.
To address this issue, the foundation model paradigm proposes a framework that transfers knowledge from a centralized foundation model for downstream use cases~\cite{foundation-model} (Fig~\ref{fig:speech_foundation_model}).
Scaling up the foundation model with more data\footnote{Either the unlabeled~\cite{oquab2023dinov2} or weakly labeled~\cite{radford2021learning} web-scale data.} and parameters improves performance on numerous downstream tasks simultaneously. This advantage has been witnessed in Natural Language Processing (NLP)~\cite{liu2019roberta} and Computer Vision (CV)~\cite{oquab2023dinov2}.
The paradigm is desirable since not all researchers have the resources to train large models from scratch for each task of interest, while transferring knowledge from a foundation model\footnote{Usually open-sourced by large corporations and publicly available.} requires minimal computational and annotation effort~\cite{bert}.
Self-Supervised Learning (SSL) has emerged as a promising technique for developing foundation models~\cite{bert,simclr}. This technique pre-trains a model with a substantial number of parameters and unlabeled data to learn powerful and transferable representations.
The pre-trained model achieves state-of-the-art (SOTA) downstream performances after fine-tuned on various tasks.
SSL appears as a realization of the foundation model paradigm to democratize SOTA deep learning research and deployment.

SSL has been explored in speech~\cite{apc1,mockingjay,tera,ling2020decoar,cpc,wav2vec,vq_wav2vec,wav2vec2,pase+,hsu2021hubert,wavlm}, with studies applying SSL models to different applications~\cite{sv-wav2vec2,lai2020semi,lin2020fragmentvc,pepino21_interspeech}. However, these studies used different downstream evaluation datasets and setups\footnote{For example, wav2vec2 was entirely finetuned for LibriSpeech ASR; TERA was evaluated with the DNN-HMM pipeline in the Pytorch-Kaldi toolkit; APC was probed for its linear separability on phonemes. Without the standardization of the evaluation protocol, it is hard to know which model is more effective for the downstream task practitioners.}. Furthermore, unlike NLP, where foundation models are assessed across multiple tasks and benchmarks like GLUE~\cite{glue,sarlin2020superglue}, speech SSL evaluation often narrows down to specific tasks (i.e. ASR\footnote{For example, wav2vec 2.0~\cite{wav2vec2}, HuBERT~\cite{hsu2021hubert} and data2vec~\cite{data2vec} are all evaluated only with ASR in their original papers.}).
Despite this approach pushing the limits for specific tasks, the approach overlooks SSL's potential for generalizing to new tasks and it remains unknown whether the techniques can lead to a foundation model for speech processing.

We introduce Speech processing Universal PERformance Benchmark (SUPERB) to study the above research question.
SUPERB standardizes speech SSL evaluation with a broad range of 15 speech processing tasks.
Compared with the traditional evaluation protocols~\cite{zerospeech,apc1} or existing SSL benchmarks~\cite{lebenchmark,parcollet2024lebenchmark,shon2022slue,shon2022slue_phase2,conneau22_interspeech,conneau2023fleurs}, SUPERB emphasizes the direct usability of SSL models on a wide spectrum of real applications.
15 tasks are selected, including phoneme recognition (PR), keyword spotting (KS), speaker identification (SID), emotion recognition (ER), intent classification (IC), slot filling (SF), automatic speech recognition (ASR), speaker verification (SV), speaker diarization (SD), query-by-example spoken term detection (QbE), speech translation (ST), out-of-domain ASR (OOD-ASR), source separation (SS), speech enhancement (SE), and voice conversion (VC).

We research SSL models' generalizability on these 15 tasks with a unified framework, which demonstrates competitive performances across tasks and can be easily extended to more tasks, as shown by Fig~\ref{fig:speech_foundation_model}.
In our framework, a lightweight prediction head is mounted on the frozen speech foundation model for each task, using weighted representations from all frozen layers. These weights are learned jointly with each task's prediction head.

Our results validate that SSL techniques are promising for building speech foundation models.
In these SSL models, distinct layers handle specific tasks, and the task-specific learnable weighted-sum enables them on various downstream tasks.
Notably, scaling up the foundation model consistently demonstrates improvements across all tasks, and top SSL models often match or exceed the performance of traditional non-SSL approaches. These findings align with those in NLP~\cite{tenney2019bert,liu2019roberta}, with the notable difference that we can achieve each task’s SOTA performance using a simple weighted-sum protocol without fine-tuning the entire speech encoder.

We defined the standardized task design, provided the baseline model results, and released the offline evaluation software in \cite{superb,tsai2022superb}.
In this work, we extend our previous studies with the following contributions:
\begin{itemize}
    \item Combined with the released evaluation codebase\footnote{\href{https://github.com/s3prl/s3prl/blob/main/s3prl/downstream/docs/superb.md}{https://github.com/s3prl/s3prl/blob/main/s3prl/downstream/docs/superb.md}}, we provide a complete platform featuring an online leaderboard supporting submissions\footnote{\href{https://superbbenchmark.org/leaderboard}{https://superbbenchmark.org/leaderboard}}. After launching the submission system, we \textit{received 14 new model submissions}, suggesting that the platform is becoming an active community. Consequently, we scale the evaluation from the original 14 models~\cite{superb} to 33 models, providing a broad coverage for the existing speech SSL literature and track the latest research (Section~\ref{section:full_benchmark_result}).
    \item We validate that SSL techniques are universal on SUPERB tasks\footnote{
    Note that the examined SSL models are primarily pre-trained on single-channel, single-speaker, read, and clean speech for English audio books (LibriSpeech~\cite{librispeech}), lacking rich variations in speaker identity, prosody, or overlapping speech. Consequently, our evaluation setting is slightly biased towards the pre-training data distribution to practically assess the SSL models’ task generalizability beyond ASR as the first step. The current SSL models are far from universal for general speech understanding and generation. A simple counterexample is their failure to comprehend spatial information due to the single-channel nature~\cite{huang2024unix,huang2024dynamic}.
    }, evidenced across various learning objectives, model configurations and computing budgets.
    The leading SSL models demonstrate strong task generalizability, achieving performances that are near or better than those of non-SSL approaches (Section~\ref{section:leading_SSL_model}), albeit except for the generation tasks that require low-level acoustic details (Section~\ref{section:generation_tasks_fail}).
    \item We observe that the learnable weighted-sum over the frozen layers of the SSL model is better than the conventional evaluation protocol: using the frozen last layer. Furthermore, individual single-layer benchmarking can sometimes yield even better results. As a result, it is desirable to sweep over all the layers to find the best layer given a specific task (Section~\ref{section:pushing_limits}). The phenomenon is especially evident on VC, since the task favors the representation with better \textit{source} speaker invariance\footnote{
    \label{footnote:speaker_invariance}
    An abundance of VC works~\cite{hsu2016voice,hsu2017voice,qian2019autovc} aim to learn speaker-independent speech representation to facilitate transferring to unseen source speakers during the testing stage. We use the term \textit{speaker invariance} to describe the degree that the representation is independent of the speaker variations.
    }. We suggest to conduct layer-wise single-layer benchmarking on VC based on this finding (Section~\ref{section:speaker_independent}).
    \item Based on the prior work~\cite{pasad2021layer}, we confirm that the layer-weights\footnote{\label{footnote:layer_weight}
    After training a weighted-sum over all the layers along with the downstream model, we acquire a weight for each layer contributing to the best performance on the development set. We term these trained weights as \textit{layer-weights} in this paper.
    } learned by the weighted-sum protocol do not reflect the layer performance precisely across SUPERB tasks. The result suggests that layer-weights are unreliable for interpreting the information flow inside an SSL model. Also, the largest layer-weight does not always relate to the best layer performance (Section~\ref{section:layer-weights-analysis}).
    \item We observe there are insignificant results between model comparisons, which potentially leads to unreliable interpretation on rankings. We then suggest to conduct statistical test when comparing to our baseline numbers. The recipes will be released (Section~\ref{section:significance}).

\end{itemize}

\section{Related Work}

Multiple benchmarks for evaluating SSL models on distinct aspect of speech have been proposed.
The ZeroSpeech series~\cite{zerospeech,dunbar2020zero} focuses on the intrinsic evaluation for different levels of content information, from phonetics, lexicon, to semantics, with the linguistically-motivated ABX-based metrics.
The SLUE series~\cite{shon2022slue,shon2022slue_phase2} benchmark SSL models for their spoken language understanding (SLU) ability like named entity recognition, sentiment analysis, and spoken question answering.
~\cite{shor2022universal} proposed a benchmark for evaluating the paralinguistic information, including masked speech detection and dysarthria classification.
In addition to the aforementioned benchmarks, there have been endeavors to evaluate various facets of speech within a unified benchmark.
\cite{wang2021fine} proposed to benchmark SSL models with SV, ER, and SLU tasks in English through fine-tuning the entire SSL model.
LeBenchmark~\cite{lebenchmark,parcollet2024lebenchmark} setups a multi-task SSL benchmark for French.
FLEURS~\cite{fleurs} and XTREME-S~\cite{conneau2022xtreme} extend the multi-task evaluation frameworks to the multi-lingual setting.

Compared to these efforts, SUPERB covers broader aspects of speech processing, including content~(ASR), semantics~(ST), speaker~(SV), prosody~(ER), denoising~(SE), conversion~(VC) and generation~(SS).
The original SUPERB~\cite{superb} benchmark addresses 10 discriminative tasks, with the follow-up SUPERB-SG~\cite{tsai2022superb} introducing 5 additional tasks for semantic and generative capabilities.
These 15 tasks define the \textit{public benchmark set} of SUPERB.
The SUPERB Challenge~\cite{feng2023superb} introduces the the concept of a \textit{hidden benchmark set} for partial tasks to prevent overfitting SSL development on the public set. The corpora for the hidden set are privately recorded and the participants submit the models to the hidden set committee for evaluation.
ML-SUPERB~\cite{shi2023ml} extends the framework to cover 143 languages in a multilingual setting, including ASR and language identification (LID) as the initial step.
In SUPERB~\cite{superb} and SUPERB-SG~\cite{tsai2022superb}, we presents the standardized task design and the evaluation results on limited models without detailed analyses and suggestions for the benchmark adoption.
We scale-up the evaluation to more up-to-date SSL models and conduct analyses to understand the best practice in this work.

\section{Speech Processing Universal PERformance Benchmark}

This section presents our unified framework for evaluating speech foundation models across numerous tasks, followed by an introduction to the tasks selected by SUPERB for benchmarking.

\subsection{Unified framework design}
\label{section:unified_training_framework}

In SUPERB, we aim to assess the task generalizability of speech foundation models fairly. This requires defining a \textit{standard interface} for evaluation and maintaining consistency in the downstream training pipeline across all models. This approach ensures that improvements in downstream tasks reflect enhancements in the foundation model itself, independent of varying downstream fine-tuning protocols.

To illustrate our unified framework, we start by defining the notations.
Given an input waveform with $S$ samples: $\bm{x}=x_1, ..., x_S \in \mathbb{R}$, the speech foundation model processes it into $L$ layers of hidden states: $\bm{h}^l=\bm{h}^l_1, ..., \bm{h}^l_{T} \in \mathbb{R}^{d}$, where $1 \leq l \leq L \in \mathbb{R}$, $T < S$, and $d \in \mathbb{R}$ is the dimension of the hidden state. It is natural to assume $d$ is consistent across all the layers since most of speech foundation model in literature adopts the same dimension across layers.

Compared to conventional evaluation protocol \cite{apc1,mockingjay} which extracts the frozen last layer $\bm{h}^L$ as the representation for downstream tasks, we consider all the layers $\bm{h}^1, ..., \bm{h}^L$ of hidden states as a single frozen representation and evaluate its quality on various speech processing tasks.
The design choice is driven by the varying types of information across different layers of BERT~\cite{tenney2019bert} in NLP. Extracting only the last layer could overlook the foundational model's capability to solve tasks that require information from earlier layers.
As a result, we define the multiple layers of hidden states $\bm{h}^1, ..., \bm{h}^L$ as our \textit{standard interface} for speech foundation models.
Any model extracting representations into this form can be considered as a candidate foundation model in our evaluation framework.

To leverage all the layers for various downstream tasks, we adopt the learnable weighted-sum approach~\cite{elmo} to reduce all layers of representation into a single representation $\bm{\tilde{h}}$:

\begin{equation}
    \bm{\tilde{h}} = \bm{\tilde{h}}_1, ..., \bm{\tilde{h}}_{T}
\end{equation}
\begin{equation}
    \bm{\tilde{h}}_t = \sum_{l=1}^{L} \alpha^l \cdot \bm{h}^l_t
\end{equation}
\begin{equation}
    \sum_{l=1}^{L} \alpha^l = 1
\end{equation}

where $1 \leq t \leq T$, $\bm{\tilde{h}}_t \in \mathbb{R}^{d}$ and $\alpha^l \geq 0$. The weights for each layer $\alpha^1, ..., \alpha^L$ are termed \textit{layer-weights} in this paper which is a valid probabilistic distribution. The learnable weighted-sum is considered as part of the downstream model. Hence, the downstream model takes $\bm{h}^1, ..., \bm{h}^L$ as the input feature, reduces it with a set of task-specific trainable weights $\alpha^1, ..., \alpha^L$, and feed the reduced representation $\bm{\tilde{h}}$ into the task-specific model.
Note that layer-weights are task-specific.
That is, there are a set of learnable weights for ASR and another set of learnable weights for SV. The layer-weights are learned jointly with the downstream ASR or SV model with gradient descent.
The design of task-specific layer-weights is motivated by our hypothesis that different tasks might favor different layers of representation, which will be verified in Section~\ref{section:information_flow}.

Throughout our downstream training process, the speech foundation model is frozen, and only the layer-weights and the downstream model parameters are optimized.
Compared to another common evaluation protocol of fine-tuning the entire foundation model as done in \cite{bert,wav2vec2,hsu2021hubert}, our design choice is made primarily due to a practical reason: the computational cost.
Fine-tuning wav2vec 2.0 Large following the official recipe in Fairseq~\cite{ott2019fairseq} takes around 4$\sim$8 GPUs which is costly for researchers in academia.
SUPERB benchmark's broad task coverage aggravates the situation, as little researchers are affordable to tune 15 tasks with 8 GPUs for each task, and this computing barrier will inevitably hinder the benchmark adoption which essentially contradicts our motivation to standardize the foundation model evaluation. 
In practice, our recipes require only a single GPU to achieve the reasonable batch size (i.e. 32) even for the Large models on all the tasks we consider.


Another important factor to consider is the source of the speech foundation models.
In principle, SUPERB does not impose restrictions on the approaches used to derive the speech foundation model.
For example, supervised~\cite{radford2023robust}, semi-supervised~\cite{xu2020iterative}, and self-supervised learning~\cite{wav2vec2,hsu2021hubert} are all possible approaches.
However, existing supervised and semi-supervised models are task-specific, and their internal representations are not easily transferable to unseen tasks~\cite{chemudupati2023transferability}.
Therefore, our focus is on exploring and standardizing the evaluation of speech foundation models using speech SSL models.

\subsection{SUPERB task design}

\begin{table*}[th]
    \centering
    \caption{
    Details of the data statistics of each adopted corpus. \textbf{Num} is the number of recordings; \textbf{Avg}, \textbf{Max} and \textbf{Min} are the average, max, and min length (in seconds) of the recordings respectively; \textbf{hour} is the total recording hours of the corpus. For QbE, the audio document database statistics is listed as the training set. The last column shows the current best method without SSL techniques on each corpus.
    }
    \resizebox{1.0\textwidth}{!}{
    \begin{tabular}{@{}cccccccccccccccccc@{}}
        \toprule
        \multirow{2}{*}[-1em]{Task} & \multirow{2}{*}[-1em]{Adopted Corpus} & \multicolumn{5}{c}{Train} & \multicolumn{5}{c}{Valid} & \multicolumn{5}{c}{Test} & \multirow{2}{*}[-1em]{\makecell{Non-SSL\\SOTA}} \\
        \cmidrule(lr){3-7} \cmidrule(lr){8-12} \cmidrule(lr){13-17} \\
        {} & {} & Num & Avg & Max & Min & Hour & Num & Avg & Max & Min & Hour & Num & Avg & Max & Min & Hour \\
        \midrule
        \midrule

PR & LibriSpeech 100 hour~\cite{librispeech} & 28539 & 12.69 & 24.52 & 1.41 & 100.59 & 2703 & 7.18 & 32.65 & 1.45 & 5.39 & 2620 & 7.42 & 34.96 & 1.28 & 5.40 & - \\
\midrule
SID & VoxCeleb1~\cite{voxceleb1} & 138361 & 8.28 & 144.92 & 3.96 & 318.41 & 6904 & 7.92 & 74.96 & 3.96 & 15.18 & 8251 & 7.86 & 76.44 & 3.96 & 18.01 & \cite{hajibabaei2018unified} \\
\midrule
\multirow{5}{*}{ER} & IEMOCAP Fold 1~\cite{iemocap} & 3556 & 4.52 & 34.14 & 0.58 & 4.47 & 890 & 4.44 & 31.91 & 0.73 & 1.10 & 1085 & 4.73 & 29.13 & 0.84 & 1.43 & \multirow{5}{*}{\cite{sarma2018emotion}} \\
{} & IEMOCAP Fold 2~\cite{iemocap} & 3606 & 4.51 & 34.14 & 0.58 & 4.52 & 902 & 4.55 & 20.29 & 0.73 & 1.14 & 1023 & 4.68 & 22.52 & 0.73 & 1.33 \\
{} & IEMOCAP Fold 3~\cite{iemocap} & 3504 & 4.61 & 34.14 & 0.73 & 4.49 & 876 & 4.43 & 29.05 & 0.76 & 1.08 & 1151 & 4.44 & 19.84 & 0.58 & 1.42 \\
{} & IEMOCAP Fold 4~\cite{iemocap} & 3600 & 4.58 & 34.14 & 0.58 & 4.58 & 900 & 4.58 & 31.91 & 0.76 & 1.14 & 1031 & 4.43 & 24.36 & 0.76 & 1.27 \\
{} & IEMOCAP Fold 5~\cite{iemocap} & 3432 & 4.56 & 24.36 & 0.58 & 4.34 & 858 & 4.62 & 29.13 & 0.73 & 1.10 & 1241 & 4.48 & 34.14 & 0.78 & 1.55 \\
\midrule
KS & Speech Commands~\cite{speech_commands} v1.0 & 51094 & 0.99 & 95.18 & 0.37 & 14.07 & 6804 & 1.04 & 95.18 & 0.38 & 1.97 & 3081 & 1.0 & 1.0 & 1.0 & 0.86 & \cite{vygon2021learning} \\
\midrule
IC & Fluent Speech Commands~\cite{fluent} & 23132 & 2.29 & 13.23 & 0.65 & 14.72 & 3118 & 2.25 & 8.36 & 0.94 & 1.95 & 3793 & 2.45 & 5.29 & 0.68 & 2.58 & \cite{fluent} \\
\midrule
ASR & LibriSpeech 100 hour~\cite{librispeech} & 28539 & 12.69 & 24.52 & 1.41 & 100.59 & 2703 & 7.18 & 32.65 & 1.45 & 5.39 & 2620 & 7.42 & 34.96 & 1.28 & 5.40 & \cite{likhomanenko2020slimipl} \\
\midrule
ASV & VoxCeleb1~\cite{voxceleb1} & 148642 & 8.24 & 144.92 & 3.96 & 340.39 & - & - & - & - & - & 4874 & 8.28 & 69.04 & 3.96 & 11.20 & \cite{okabe2018attentive} \\
\midrule
SD & Libri2Mix (noisy, max)~\cite{cosentino2020librimix} & 13900 & 14.59 & 24.52 & 3.22 & 56.37 & 3000 & 9.12 & 28.57 & 3.08 & 7.60 & 3000 & 8.41 & 21.25 & 3.09 & 7.01 & - \\
\midrule
QbE & Quesst14 (English)~\cite{quesst2014} & 2438 & 6.72 & 33.47 & 3.01 & 4.55 & 138 & 1.47 & 6.19 & 1.02 & 0.06 & 138 & 1.44 & 2.79 & 1.01 & 0.06 & \cite{gtts} \\
\midrule
SF & Audio SNIPS~\cite{coucke2018snips,lai2021semi} & 104672 & 2.85 & 10.71 & 0.57 & 82.93 & 2800 & 2.94 & 8.59 & 0.91 & 2.29 & 2800 & 2.89 & 7.29 & 0.99 & 2.25 & \cite{chen2019bert} \\
\midrule
\multirow{5}{*}[0.5em]{OOD-ASR} & Common Voice 7.0 (es)~\cite{common-voice} & 13756 & 5.61 & 10.80 & 1.78 & 21.44 & 733 & 5.84 & 10.39 & 2.28 & 1.19 & 366 & 6.12 & 11.42 & 2.40 & 0.62 & \multirow{5}{*}[0.5em]{\cite{radford2023robust}} \\
{} & Common Voice 7.0 (zh)~\cite{common-voice} & 21266 & 5.25 & 19.56 & 1.49 & 31.04 & 9334 & 5.56 & 10.99 & 1.70 & 14.41 & 9338 & 5.91 & 11.74 & 1.73 & 15.32 \\
{} & Common Voice 7.0 (ar)~\cite{common-voice} & 27168 & 4.03 & 22.03 & 1.92 & 30.38 & 10144 & 4.34 & 10.51 & 1.40 & 12.23 & 10271 & 4.37 & 10.44 & 1.56 & 12.46 \\
{} & SBCSAE (spon)~\cite{du2000sbcsae} & 30339 & 1.35 & 14.83 & 0.10 & 11.42 & 4646 & 1.23 & 13.03 & 0.10 & 1.59 & 5010 & 1.55 & 12.78 & 0.15 & 2.15 \\
\midrule
ST & CoVoST2 (En to De)~\cite{covost2} & 288187 & 5.34 & 24.67 & 0.98 & 427.72 & 15480 & 6.05 & 30.26 & 1.54 & 26.02 & 15507 & 5.71 & 142.54 & 1.10 & 24.61 & \cite{ST-SOTA} \\
\midrule
\multirow{4}{*}{VC} & VCC2020 Task1 (TEF1)~\cite{vcc2020} & 60 & 3.16 & 4.26 & 1.29 & 0.05 & \multirow{4}{*}{-} & \multirow{4}{*}{-} & \multirow{4}{*}{-} & \multirow{4}{*}{-} & \multirow{4}{*}{-} & \multirow{4}{*}{100} & \multirow{4}{*}{2.99} & \multirow{4}{*}{5.05} & \multirow{4}{*}{1.34} & \multirow{4}{*}{0.08} & \multirow{4}{*}{\cite{zhang2020voice}} \\
{} & VCC2020 Task1 (TEF2)~\cite{vcc2020} & 60 & 3.62 & 5.80 & 1.76 & 0.06 & {} & {} & {} & {} & {} & {} & {} & {} & {} & {} \\
{} & VCC2020 Task1 (TEM1)~\cite{vcc2020} & 60 & 4.29 & 6.21 & 1.63 & 0.07 & {} & {} & {} & {} & {} & {} & {} & {} & {} & {} \\
{} & VCC2020 Task1 (TEM2)~\cite{vcc2020} & 60 & 3.75 & 5.49 & 1.79 & 0.06 & {} & {} & {} & {} & {} & {} & {} & {} & {} & {} \\
\midrule
SE & Voicebank-DEMAND~\cite{veaux2013voice} & 10802 & 2.92 & 15.11 & 1.09 & 8.76 & 770 & 2.96 & 11.78 & 1.36 & 0.63 & 824 & 2.51 & 9.77 & 1.24 & 0.58 & \cite{abdulatif2022cmgan} \\
\midrule
SS & Libri2Mix (clean, min)~\cite{cosentino2020librimix} & 13900 & 11.21 & 16.60 & 3.0 & 43.27 & 3000 & 5.42 & 17.47 & 3.01 & 4.51 & 3000 & 5.02 & 13.99 & 3.0 & 4.19 & \cite{li2022efficient} \\

        \bottomrule
    \end{tabular}}
    \label{table:statistics}
\end{table*}

SUPERB is designed with the following principles:
\begin{enumerate}
    \item \textbf{Task generalizability}: 
    SUPERB standardizes the comparison of SSL models across 15 diverse speech processing tasks, covering content, speaker characteristics, prosody, semantics, and generation. Unlike previous benchmarks~\cite{zerospeech,lebenchmark}, SUPERB uniquely emphasizes task generalizability, making it the benchmark with the most comprehensive coverage of diverse speech processing tasks\footnote{
    In SUPERB, we evaluate the usability of SSL representations on a series of new tasks. These tasks and their annotations are unseen during the SSL pre-training phase. The broad and diverse task coverage ensures that we derive a reliable conclusion regarding the representations’ \textit{task generalizability}.}.
    \item \textbf{Community standard}: 
    SUPERB incorporates tasks from speech communities and adheres to conventional evaluation protocols to align with common research interests. Unlike previous approaches that focus on linear separability~\cite{apc1,mockingjay} or intrinsic properties~\cite{zerospeech}, such as ABX score, we focus on the direct usability of SSL models on real applications, connecting to research beyond representation learning.
    \item \textbf{Open access}: SUPERB is open-sourced, with all materials publicly accessible. We select corpora with open licenses and release our evaluation codebase, including all data pre-processing steps, to ensure reproducibility.
\end{enumerate}

To meet these principles, a broad variety of speech tasks were chosen based on the most popular tasks on which results were reported at Interspeech 2020\footnote{\href{http://www.interspeech2020.org/Program/Technical_Program/}{http://www.interspeech2020.org/Program/Technical\_Program/}}.
15 tasks covering 5 dimensions are included:

\begin{enumerate}
    \item \textbf{Content}: Phoneme Recognition~(PR), Automatic Speech Recognition~(ASR), Out-of-domain ASR~(OOD-ASR), Keyword Spotting~(KS), Query-by-example~(QbE)
    \item \textbf{Speaker}: Speaker Identification~(SID), Speaker Verification~(SV), Speaker Diarization~(SD)
    \item \textbf{Prosody}: Emotion Recognition~(ER)
    \item \textbf{Semantics}: Intent Classification~(IC), Slot Filling~(SF), Speech Translation~(ST)
    \item \textbf{Generation}: Voice Conversion~(VC), Source Separation~(SS), and Speech Enhancement~(SE)
\end{enumerate}

In the following, we describe each task design in detail, and we list the data statistics and the non-SSL baseline method for each corpus in Table~\ref{table:statistics}.
Note that we do not compare to the results of fine-tuning the SSL encoder since the approach usually yields better results but is computational costly and not considered in our framework.
Two tasks do not have prior works: PR and SD.
We illustrate the reasons in the following task sections.
In each task, we describe the goal of the task, its realization corpus we select, and the evaluation protocol we follow including the data splits and evaluation metrics.
Furthermore, we illustrate how we leverage speech SSL model to solve the task, involving the downstream model architecture and the optimization loss.

\subsubsection{\textbf{Phoneme Recognition (PR)}}

\begin{itemize}
    \item \textit{Task description:} PR transcribes an utterance into the smallest spoken unit: phoneme.
We include alignment modeling in the PR task to avoid the potential inaccurate forced alignment.
During the downstream model training, the labeled data for each utterance is the raw phoneme sequence without the boundary information, and the downstream model needs to learn the alignment relying on the foundation model representation.
    \item \textit{Data and metrics:} 
    Phoneme recognition is commonly conducted on the TIMIT~\cite{timit} dataset. However, since TIMIT is not freely available which violates our \textit{Open access} principle, we opt for LibriSpeech for this task and thus no prior result is comparable.
    LibriSpeech~\cite{librispeech} train-clean-100/dev-clean/test-clean subsets are adopted for training/validation/testing. Phoneme transcriptions are obtained from the official lexicon file\footnote{\href{https://www.openslr.org/11/}{https://www.openslr.org/11/}}.
    The \textit{g2p-model-5} and the conversion script in the Kaldi \textit{librispeech s5}\footnote{\href{https://github.com/kaldi-asr/kaldi/tree/master/egs/librispeech/s5}{https://github.com/kaldi-asr/kaldi/tree/master/egs/librispeech/s5}} recipe are used when an out-of-vocabulary (OOV) word is encountered.
    The evaluation metric is phone error rate (PER).
    \item \textit{Downstream model:} We train a frame-wise 2-layer linear model as the downstream model on top of the foundation model representation. The downstream model is optimized by the CTC~\cite{ctc} loss.
\end{itemize}

\subsubsection{\textbf{Automatic Speech Recognition (ASR)}}

\begin{itemize}
    \item \textit{Task description:} ASR transcribes utterances directly into words. Compared to PR, ASR further involves converting the spoken units into the writing units. While PR analyzes the improvement in modeling phonetics, ASR reflects the significance of the improvement in a more real-world scenario.
    \item \textit{Data and metrics:} LibriSpeech train-clean-100/dev-clean/test-clean subsets are used for training/validation/testing. The evaluation metric is word error rate (WER).
    \item \textit{Downstream model:} A vanilla 2-layer 1024-unit BLSTM is adopted as the downstream model and optimized by CTC loss on characters. SpecAugment~\cite{specaug} is also applied to the representations to avoid overfitting.
\end{itemize}

\subsubsection{\textbf{Out-Of-Domain ASR (OOD-ASR)}}

\begin{itemize}
    \item \textit{Task description:}
The vanilla ASR task only examines foundation models' ability on the read English corpus LibriSpeech~\cite{librispeech} which does not involve speaking-style variations.
Also, most of the SSL models use LibriSpeech as the pre-training data.
Hence, PR and ASR are the in-domain downstream tasks.
We consider the out-of-domain scenarios across languages and speaker styles.
    \item \textit{Data and metrics:}
The OOD-ASR tasks are categorized into \textit{cross-lingual} and \textit{spontaneous speech} tasks.
For the cross-lingual tasks, we choose the Mexican Spanish (es), Mandarin (zh), and Arabic (ar) subsets from Common Voice 7.0~\cite{common-voice}.
For the spontaneous speech task (spon), we use the Santa Barbara Corpus of Spoken American English (SBCSAE)~\cite{du2000sbcsae}, consisting of 60 conversations over different topics\footnote{The data pre-processing follows \url{https://github.com/vectominist/SBCSAE-preprocess}}.
The standard split from each corpus is adopted.
We WER as the metric except for Mandarin which character error rate (CER) is used. The error rates are averaged across 4 sub-tasks to offer an overall OOD-ASR score.
    \item \textit{Downstream model:}
The OOD-ASR task shares the same downstream model with the ASR task, including the model architecture and the optimization loss.
\end{itemize}

\subsubsection{\textbf{Keyword Spotting (KS)}}

\begin{itemize}
    \item \textit{Task description:}
Compared to the content recognition tasks listed above, the content detection tasks involve detecting the pre-registered spoken terms.
A KS system detects pre-registered keywords and ignore the un-registered words.
For example, the system should be awaken by the "Hey Siri" spoken command and remain silence when hearing the irrelevant content.
We approach this task by classifying an input utterance into pre-defined \textit{keyword} classes and an \textit{unknown} class for the un-registered words.
    \item \textit{Data and metrics:}
Speech Commands dataset v1.0~\cite{speech_commands} is used.
The dataset consists of ten classes of keywords, a class for silence, and an \textit{unknown} class.
The standard split of the corpus is adopted.
The evaluation metric is accuracy (ACC).
    \item \textit{Downstream model:}
A simple linear model followed by the mean pooling is used as the downstream model and trained with the cross entropy loss.
\end{itemize}

\subsubsection{\textbf{Query by Example Spoken Term Detection (QbE)}}

\begin{itemize}
    \item \textit{Task description:}
QbE is another content detection task. It detects a spoken term (short query) in an audio database (long documents) by binary discriminating whether the query appears in each document.
In practice, given a spoken query, the QbE system assigns a continuous matching score to each of the spoken document.
    \item \textit{Data and metrics:}
The English subset\footnote{The original corpus is multilingual.} in QUESST 2014~\cite{quesst2014} challenge is adopted.
The corpus is composed of three parts: spoken documents, development spoken queries and testing spoken queries.
The development and testing queries share the same set of the spoken documents.
Each query is labeled by the ids of the documents containing the corresponding spoken term.
The evaluation metric is maximum term weighted value (MTWV)\footnote{\href{https://www.nist.gov/system/files/documents/itl/iad/mig/OpenKWS13-EvalPlan.pdf}{https://www.nist.gov/system/files/documents/itl/iad/mig/OpenKWS13-EvalPlan.pdf}}.
    \item \textit{Downstream model:}
We follow the system proposed by GTTS-EHU for QUESST at MediaEval 2014 \cite{gttsehu} but replace the conventional supervised phoneme posteriorgram (PPG) with SSL representations.
Specifically, we run Dynamic Time Warping\cite{dtw}~(DTW) to obtain a similarity score for each query-document pair.
The scores belonging to the same query are further normalized.
The hyper-parameters of the DTW include the distance function used for measuring the query-document similarity, and the layer we extract from the speech foundation model.
The best pair of distance function and the layer id found on the development set (queries) is used to report the performance on the test set (queries).

\end{itemize}

\subsubsection{\textbf{Speaker Identification (SID)}}

\begin{itemize}
    \item \textit{Task description:}
SID recognizes each utterance for its speaker identity as a multi-class classification, where speakers are in the same pre-defined set for both training and testing.
    \item \textit{Data and metrics:}
The widely used VoxCeleb1~\cite{voxceleb1} is adopted following the standard split for the classification task.
The evaluation metric is accuracy (ACC).
    \item \textit{Downstream model:}
A simple linear head followed by the mean pooling is used as the downstream model and trained with the cross entropy loss.
\end{itemize}

\subsubsection{\textbf{Speaker Verification (SV)}}

\begin{itemize}
    \item \textit{Task description:}
SV involves determining whether the speakers of two utterances (enrollment and testing) are the same, functioning as a binary classification task. Unlike SID, SV poses a greater challenge because the speakers in the testing set may not be present in the training set. This aspect aligns SV more closely with real-world speaker authentication systems, which often encounter speakers not previously encountered by the system.
    \item \textit{Data and metrics:}
VoxCeleb1~\cite{voxceleb1} is used for training without VoxCeleb2 training data and MUSAN~\cite{snyder2015musan} noise augmentation.
The standard testing set and the testing pairs are adopted.
The evaluation metric we use is equal error rate (EER).
    \item \textit{Downstream model:}
We adopt the well-known x-vector~\cite{snyder2018x} as the downstream model.
The model is trained on VoxCeleb1 with the AMSoftmax loss~\cite{wang2018additive} following the hyper-parameters described in \cite{voxceleb1}.
After we train the classification model, the learned hidden state is used as the speaker embedding.
We compute cosine-similarity between the speaker embeddings for each pairs of the enrollment and the testing utterances to produce a matching score.
Finally, the binary decision threshold is determined when computing EER.
\end{itemize}

\subsubsection{\textbf{Speaker Diarization (SD)}}

\begin{itemize}
    \item \textit{Task description:}
SD predicts \textit{who is speaking when}. 
Compared to SID and SV, SD requires the speaker information for each distinct timestamp in a conversation.
Furthermore, multiple speakers can speak simultaneously.
Thus, the speech foundation model has to encode rich speaker characteristics for each timestamp and should be able to represent mixtures of signals.
    \item \textit{Data and metrics:}
We curate the speaker diarization labels from LibriMix, hence no prior result is available.
LibriMix~\cite{cosentino2020librimix} is derived from LibriSpeech where train-clean-100/dev-clean/test-clean are used to generate mixtures for training/validation/testing.
We us the Libri2Mix subset.
We employ 100-hour clean-train, clean-dev and clean-test to generate training, development, and test mixtures, respectively.
The WHAM!~\cite{wichern2019wham} noises are augmented to the utterances.
The time-coded speaker labels were generated using alignments from Kaldi LibriSpeech ASR model.
The evaluation metric is diarization error rate (DER).
    \item \textit{Downstream model:}
We employ the end-to-end training scheme with permutation-invariant training (PIT) loss \cite{fujita2019end} to SD, instead of using clustering-based methods.
We leverage a single-layer 512-unit LSTM as the downstream model.
\end{itemize}

\subsubsection{\textbf{Emotion Recognition (ER)}}

\begin{itemize}
    \item \textit{Task description:}
ER recognize the emotion category for the affected speech.
The task examines speech foundation model's ability of encoding the prosody information.
    \item \textit{Data and metrics:}
IEMOCAP~\cite{iemocap} is adopted, and we follow the conventional evaluation protocol: we drop the unbalance emotion classes to leave the final four classes (neutral, happy, sad, angry) with a similar amount of data points and cross-validates on five folds of the standard splits.
The evaluation metric is accuracy (ACC).
    \item \textit{Downstream model:}
A simple linear head followed by the mean pooling is used as the downstream model and trained with the cross entropy loss.
\end{itemize}

\subsubsection{\textbf{Intent Classification (IC)}}

\begin{itemize}
    \item \textit{Task description:}
As a component of spoken language understanding (SLU), IC involves recognizing spoken commands and categorizing them into predefined intent classes. Unlike conventional methods~\cite{coucke2018snips} that transcribe utterances and then interpret intent from text, our approach is end-to-end, designed to evaluate the foundation model's ability to directly understand semantic meaning.
    \item \textit{Data and metrics:}
We use the Fluent Speech Commands~\cite{lugosch2019speech} dataset, where each utterance is tagged with three intent types: action, object, and location.
The standard split is adopted.
The evaluation metric is accuracy (ACC).
Note that we only count a full match for all three intent types as a correct prediction.
    \item \textit{Downstream model:}
A simple linear head followed by the mean pooling is used as the downstream model and trained with the cross entropy loss.
\end{itemize}

\subsubsection{\textbf{Slot Filling (SF)}}

\begin{itemize}
    \item \textit{Task description:}
In a SLU system, a recognized intent is associated with a list of \textit{entities} that must be extracted from the user's query~\cite{coucke2018snips}.
For instance, in the query "Find me a flight from Paris to New York," after the SLU system identifies the intent as \textit{searchFlight}, it needs to extract Paris as the \textit{origin} and New York as the \textit{destination} for the search. In this example, \textit{origin} and \textit{destination} are \textbf{slot types}, while Paris and New York are the corresponding \textbf{slot values}.
SF then requires the model to derive all slot types and their corresponding slot values from the input utterance in an end-to-end manner.
    \item \textit{Data and metrics:}
Audio SNIPS~\cite{lai2020towards} is adopted, which synthesized multi-speaker utterances for SNIPS~\cite{coucke2018snips}.
Following the standard split in SNIPS\footnote{The original SNIPS only defines the standard split on text without the speech data. AudioSNIPS synthesized speech for each text with 16 speakers. It is important to prevent speaker-overlapping among training/validation/testing splits. Hence, further speaker partition should be decided.}, US-accent speakers are further selected for training, and others are for validation/testing.
The evaluation metrics include slot type F1 score and slot value CER~\cite{tomashenko2019recent}.
The former evaluates predicted slot types' correctness without considering slot values; the latter compute CER between the predicted and the ground-truth slot values.

    \item \textit{Downstream model:}
We represent slot types as special tokens to wrap the slot values in transcriptions.
For example, \textit{"flying from Paris to New York"} is transformed into \textit{"flying from \textless origin\textgreater\ Taipei \textless /origin\textgreater\ to\ \textless destination\textgreater\ New York\ \textless /destination\textgreater"}. 
The special tokens are in a pre-defined set, hence we can consider them as the limited new characters.
SF is then re-formulated as an ASR problem.
The downstream model is the same as in our ASR task, except for the pre-processing to encode slot types into transcriptions and post-processing to decode slot types and slot values from hypotheses.
SpecAugment~\cite{specaug} is applied as well to the representations to avoid overfitting.
\end{itemize}

\subsubsection{\textbf{Speech Translation (ST)}}

\begin{itemize}
    \item \textit{Task description:}
ST involves translating the acoustic speech signals in the source language directly into the words in the target language.
We use it to evaluate the semantic capability of speech foundation models.
    \item \textit{Data and metrics:}
We use the CoVoST2 En$\rightarrow$De~\cite{covost2} dataset with their official train, validation, and test splits while removing all the samples containing "REMOVE".
For text, we keep original case, normalize punctuation, and build character vocabulary with 100\% train-set coverage.
We report case-sensitive de-tokenized BLEU using sacreBLEU~\cite{sacrebleu}.
    \item \textit{Downstream model:}
Our downstream model is an encoder-decoder architecture with 3 layers of Transformers each with hidden dimension of 512.
A convolutional sub-sampler is used to reduce the sequence length of the input before feeding it to the encoder.
We train our model with label-smoothing using a probability of 0.1. A beam size of 20 is used for inference.
\end{itemize}

\subsubsection{\textbf{Voice Conversion (VC)}}

\begin{itemize}
    \item \textit{Task description:}
VC is a generative task involving converting the speaking styles (speaker, accent, emotion, etc) while preserving the linguistic content.
In this task, we consider converting the speaker characteristics under the any-to-one (A2O) setting.
A2O VC aims to convert speech from any unseen speaker into that of a pre-defined target speaker.
    \item \textit{Data and metrics:}
We follow the intra-lingual VC task in VCC2020~\cite{vcc2020}.
The dataset is composed of 4 source speakers and 4 target speakers in English.
Since we consider the A2O setting, we simplify the discussion to a single target speaker in the following and the procedure is repeatedly conducted for each of the 4 target speakers.
Given a target speaker, there are 60 utterances for training and no validation is used.
After the training, 25 testing utterances from each unseen source speaker are used to test the conversion.
100 conversions are tested in total.
Each converted utterance has a reference utterance, we use mel-cepstrum distortion (MCD), word error rate (WER) and automatic speaker verification accept rate (ASV-acc) from off-the-shelf ASR and ASV models as evaluation metrics.
The correlation between these objective metrics and the subjective evaluation is justified in \cite{huang2022s3prl}.
After repeating the above procedure for 4 target speakers, all 16 source-target speaker pairs are tested.
    \item \textit{Downstream model:}
We adopt the recognition-synthesis framework illustrated in \cite{huang2022s3prl}, where the recognizer should extract the linguistic content and the synthesizer generates speech in the target speaker style grounded on the recognized content.
For the recognizer, SSL models are used to replace the traditional PPG~\cite{liu2018wavenet}.
For the synthesizer, we train a Tacotron2~\cite{Taco2} to map SSL features to FBANK for each target speaker.
Four synthesizers are trained in total.
The target speaker characteristics are modeled directly by the synthesizer without requiring a target speaker embedding.
Finally, the synthesized FBANK is decoded to waveforms by the Hifi-GAN~\cite{hifigan} vocoder.
\end{itemize}

\subsubsection{\textbf{Speech Separation (SS)}}

\begin{itemize}
    \item \textit{Task description:}
SS is a generative task of separating target speech from background interference~\cite{wang2018supervised}.
It is an important step for the multi-talker scenarios.
This task is used to evaluate the capability of speech foundation models to handle mixture of acoustic signals and separate the human speech from different speakers.
    \item \textit{Data and metrics:}
Libri2Mix~\cite{cosentino2020librimix} is adopted, which is a dataset simulated from LibriSpeech.
Each sample is a two-speaker mixture.
We use the 16kHz min-clean version of the dataset.
We use the scale-invariant signal-to-distortion ratio improvement (SI-SDRi) as the evaluation metric.
    \item \textit{Downstream model:}
We use a 3-layer BLSTM as the downstream model with dimension of 896 to predict the short-time Fourier transform (STFT) masks for each speaker.
The masks are applied to the STFT of the input (mixed) utterance to construct the predicted STFT for a target speaker.
The predicted STFT are transformed back to the time domain using inverse short-time Fourier transform (iSTFT).
Permutation invariant training (PIT)~\cite{yu2017permutation} is used to optimize the mean square error between the predicted mask and Ideal Non-negative Phase Sensitive Mask (INPSM)~\cite{erdogan2015phase,kolbaek2017multitalker}. We choose frequency domain method instead of a time domain based method because of the stride size constraint and computational cost.
\end{itemize}

\subsubsection{\textbf{Speech Enhancement (SE)}}

\begin{itemize}
    \item \textit{Task description:}
SE is a generative task of removing background noise from a distorted speech signal, and it aims to improve the perceived quality and intelligibility of the signal.
    \item \textit{Data and metrics:}
We use Voicebank-DEMAND~\cite{veaux2013voice} following the standard split.
Our evaluation metrics are Perceptual Evaluation of Speech Quality (PESQ) and Short-Time Objective Intelligibility (STOI).
    \item \textit{Downstream model:}
We follow the mask-based speech enhancement pipeline in~\cite{kolbaek2017multitalker}.
A 3-layer BLSTM model similar to the SS task is adopted as the prediction head and trained to predict the spectral mask for the clean signal.
The prediction is transformed back to the time domain using inverse short-time Fourier transform (iSTFT).
The mean square error between the predicted mask and INPSM is used as the objective.
\end{itemize}

\subsection{Acknowledgement}

Our approach does not involve developing new SSL methods or new datasets. Instead, we leverage the extensive resources from the open-source community, including SSL pre-training codebases, pre-trained SSL models, and annotated datasets. Our main contributions are:
\begin{enumerate}
    \item Connecting SSL development with the conventional non-SSL approaches on various real applications.
    \item Revealing the strong task generalizability of the leading SSL models.
    \item Providing comprehensive evaluation results to standardize the evaluation of numerous speech SSL models.
    \item Open-sourcing our evaluation codebase to simplify reproducing baselines on all the tasks.
\end{enumerate}
Given our reliance on existing resources, we recommend researchers to continue using individual datasets and consider our setting as one specific use case for evaluating speech foundation models.

\begin{table*}[th]
    \centering
\caption{
Details of 33 investigated speech foundation models. LibriSpeech and LibriLight are denoted as LS and LL, respectively. LVG stands for the 94k-hour dataset consisting of LibriLight, VoxPopuli, and Gigaspeech. SC stands for SpokenCOCO.
}
    \resizebox{0.95\textwidth}{!}{
    \begin{tabular}{@{}lcccccc@{}}
        \toprule
        Model & \#Params & Corpus & Implementation (GitHub) & \makecell{Use official \\ checkpoint} & \makecell{Community \\ submission} & \makecell{Tasks underperforming\\FBANK} \\

\midrule
\midrule

FBANK & - & - & pytorch/audio~\cite{yang2022torchaudio} & - & - & - \\

\midrule

PASE+~\cite{pase+} & 7.83M & LS 50 hr & santi-pdp/pase & \checkmark & \texttimes & ASR, SF, SV, VC \\

\midrule

APC~\cite{apc1} & 4.11M & LS 360 hr & \multirow{2}{*}{\makecell{iamyuanchung/Autoregressive-Predictive-Coding \\ Alexander-H-Liu/NPC}} & \texttimes & \texttimes & SD, SS \\


VQ-APC~\cite{vq_apc} & 4.63M & LS 360 hr & {} & \texttimes & \texttimes & SF, SD, SS \\

\midrule

NPC~\cite{npc} & 19.38M & LS 360 hr & Alexander-H-Liu/NPC & \texttimes & \texttimes & SE, SS \\

\midrule

Mockingjay~\cite{mockingjay} & 85.12M & LS 360 hr & \multirow{5}{*}{s3prl/s3prl} & \checkmark & \texttimes & OOD-ASR, SF, QbE, SV, SD, SE \\


TERA~\cite{tera} & 21.33M & LS 960 hr & {} & \checkmark & \texttimes & SF, QbE, SV, SE \\

Audio Albert~\cite{audio_albert} & 7.15M & LS 960 hr & {} & \checkmark & \texttimes & QbE \\

DistilHuBERT~\cite{chang2022distilhubert} & 27.03M & LS 960 hr & {} & \checkmark & \checkmark & SS \\

\midrule

DeCoAR~\cite{decoar} & 67.25M & LS 960 hr & \multirow{2}{*}{awslabs/speech-representations} & \checkmark & \texttimes & - \\

DeCoAR 2.0~\cite{ling2020decoar} & 89.84M & LS 960 hr & {} & \checkmark & \texttimes & SE, SS \\

\midrule

Modified CPC~\cite{modified_cpc} & 1.84M & LL 60k hr & facebookresearch/CPC\_audio & \checkmark & \texttimes & SV, SD \\

\midrule

data2vec-aqc Base~\cite{data2vec-aqc} & 93.84M & LS 960 hr & Speech-Lab-IITM/data2vec-aqc & \checkmark & \checkmark & - \\ 

\midrule

CCC-wav2vec 2.0 Base~\cite{lodagala2023ccc} & 95.04M & LS 960 hr & Speech-Lab-IITM/CCC-wav2vec-2.0 & \checkmark & \checkmark & - \\

\midrule

FaST-VGS+~\cite{fast-vgs-plus} & 217.23M &  \makecell{LS 960 hr \\ SC 742 hr} & jasonppy/FaST-VGS-Family & \checkmark & \checkmark & - \\

\midrule

LightHuBERT Stage1~\cite{lighthubert} & 94.38M & LS 960 hr & \multirow{2}{*}{mechanicalsea/lighthubert} & \checkmark & \checkmark & - \\

LightHuBERT Small~\cite{lighthubert} & 26.88M & LS 960 hr & {} & \checkmark & \checkmark & SE \\

\midrule

CoBERT Base~\cite{cobert} & 94.35M & LS 960 hr & mct10/CoBERT & \checkmark & \checkmark & SE, SS \\

\midrule

wav2vec~\cite{wav2vec} & 32.54M & LS 960 hr & \multirow{9}{*}{pytorch/fairseq~\cite{ott2019fairseq}} & \checkmark & \texttimes & SE \\

vq-wav2vec~\cite{vq_wav2vec} & 32.15M & LS 960 hr & {} & \checkmark & \texttimes & SV, SE, SS \\


wav2vec 2.0 Base~\cite{wav2vec2} & 95.04M & LS 960 hr & {} & \checkmark & \texttimes & - \\

wav2vec 2.0 Large~\cite{wav2vec2} & 317.38M & LL 60k hr & {} & \checkmark & \texttimes & SE \\

HuBERT Base~\cite{hsu2021hubert} & 94.68M & LS 960 hr & {} & \checkmark & \texttimes & - \\

HuBERT Large~\cite{hsu2021hubert} & 316.61M & LL 60k hr & {} & \checkmark & \texttimes & - \\

Data2vec Base~\cite{data2vec} & 93.16M & LS 960 hr & {} & \checkmark & \checkmark & SE, SS \\

Data2vec Large~\cite{data2vec} & 313.28M & LL 60k hr & {} & \checkmark & \checkmark & - \\

\midrule

DPHuBERT~\cite{peng2023dphubert} & 23.59M & LS 960hr & \multirow{2}{*}{pyf98/DPHuBERT} & \checkmark & \checkmark & - \\

DPWavLM~\cite{peng2023dphubert} & 23.59M & LS 960hr & {} & \checkmark & \checkmark & SS \\

\midrule

Unispeech SAT Base~\cite{unispeech-sat} & 94.37M & LS 960 hr & \multirow{7}{*}{microsoft/UniSpeech} & \checkmark & \texttimes & - \\

Unispeech SAT Base+~\cite{unispeech-sat} & 94.37M & LVG 94k hr & {} & \checkmark & \texttimes & - \\

Unispeech SAT Large~\cite{unispeech-sat} & 315.43M & LVG 94k hr & {} & \checkmark & \texttimes & - \\

WavLM Base~\cite{wavlm} & 94.38M & LS 960 hr & {} & \checkmark & \checkmark & - \\

WavLM Base+~\cite{wavlm} & 94.38M & LVG 94k hr & {} & \checkmark & \checkmark & - \\

WavLM Large~\cite{wavlm} & 315.45M & LVG 94k hr & {} & \checkmark & \checkmark & - \\

\bottomrule
    \end{tabular}}
\label{table:upstream_info}
\end{table*}

\section{Main Result}

To leverage speech SSL models, we follow the official release for model definitions, pre-trained weights, and the model forward pipelines if not mentioned specifically.
If the pre-trained weights are not available, we pre-train the model with the released codebase following the default config files.
We list all the 33 models we explored in this work in Table~\ref{table:upstream_info}.
The results obtained from the public leaderboard submissions are highlighted by the \textit{Community submission} column.

Note that it is important to explicitly search for the suitable learning rates for different SSL models instead of directly using the default one in our released codebase, since different models favor different learning rates as shown by Table~\ref{table:compare_lr}.
We search from 1e-1 to 1e-7 in log-scale in the following experiments.

\begin{table}[t!]
    \centering
    \caption{Weighted-sum benchmark results of wav2vec 2.0 and HuBERT on SID and IC using different fine-tuning learning rates. The learning rates with $\ast$ denote the default learning rate in our codebase.}
    \resizebox{0.48\textwidth}{!}{
    \begin{tabular}{lcccccc}
        \toprule
        \multirow{2}{*}[-1em]{Models} & \multicolumn{4}{c}{SID (acc)} & \multicolumn{2}{c}{IC (acc)}\\
        \cmidrule(lr){2-5} \cmidrule(lr){6-7} \\
        {} & 1e-1 & 1e-2 & 1e-3 & 1e-4\textsuperscript{$\ast$} & 1e-3 & 1e-4\textsuperscript{$\ast$} \\
        \midrule
        wav2vec 2.0 Base & NaN & 74.28 & \textbf{75.18} & 66.72 & 92.12 & \textbf{92.35} \\
        wav2vec 2.0 Large & NaN & 84.38 & \textbf{86.15} & 82.71 & \textbf{95.28} & 93.22 \\
        HuBERT Base & \textbf{81.42} & 81.01 & 70.09 & 67.37 & \textbf{98.34} & 97.81 \\
        HuBERT Large & NaN & 86.94 & \textbf{90.33} & 86.94 & 98.63 & \textbf{98.76} \\
        \bottomrule
    \end{tabular}
    }
\label{table:compare_lr}
\end{table}

\subsection{Last-layer feature vs. learnable weighted-sum}
\label{section:different_finetune_protocols}

\begin{table}[t]
    \centering
    \caption{
The last layer representation v.s. weighted-sum over all layers. In each cell, the upper number represents the last layer; the lower number represents the weighted-sum. Bold fonts highlight the cases when weighted-sum is worse.
}
    \resizebox{0.355\textwidth}{!}{
    \begin{tabular}{@{}lccccc@{}}
        \toprule
        \multirow{2}{*}[-1em]{Models} & PR & KS & SID & IC & ER \\
        \cmidrule(lr){2-2} \cmidrule(lr){3-3} \cmidrule(lr){4-4} \cmidrule(lr){5-5} \cmidrule(lr){6-6} \\
        {} & per $\downarrow$ & acc $\uparrow$ & acc $\uparrow$ & acc $\uparrow$ & acc $\uparrow$ \\
        \midrule
        \midrule

PASE+~\cite{pase+} & \makecell{58.88 \\ 58.87} & \makecell{82.37 \\ 82.54} & \makecell{35.84 \\ 37.99} & \textbf{\makecell{30.29 \\ 29.82}} & \makecell{57.64 \\ 57.86} \\

\midrule

APC~\cite{apc1} & \textbf{\makecell{41.85 \\ 41.98}} & \textbf{\makecell{91.04 \\ 91.01}} & \makecell{59.79 \\ 60.42} & \makecell{74.64 \\ 74.69} & \makecell{58.84 \\ 59.33} \\

\midrule

VQ-APC~\cite{vq_apc} & \makecell{42.86 \\ 41.08} & \makecell{90.52 \\ 91.11} & \makecell{49.57 \\ 60.15} & \makecell{70.52 \\ 74.48} & \makecell{58.31 \\ 59.66} \\

\midrule

TERA~\cite{tera} & \textbf{\makecell{47.53 \\ 49.17}} & \makecell{88.09 \\ 89.48} & \textbf{\makecell{58.67 \\ 57.57}} & \makecell{48.80 \\ 58.42} & \makecell{54.76 \\ 56.27} \\

\midrule

wav2vec~\cite{wav2vec} & \makecell{32.39 \\ 31.58} & \makecell{94.09 \\ 95.59} & \makecell{44.88 \\ 56.56} & \makecell{78.91 \\ 84.92} & \makecell{58.17 \\ 59.79} \\

\midrule

vq-wav2vec~\cite{vq_wav2vec} & \makecell{53.49 \\ 33.48} & \makecell{92.28 \\ 93.38} & \textbf{\makecell{39.04 \\ 38.80}} & \makecell{59.4 \\ 85.68} & \makecell{55.89 \\ 58.24} \\

\midrule

wav2vec 2.0 Base~\cite{wav2vec2} & \makecell{28.37 \\ 5.74} & \makecell{92.31 \\ 96.23} & \makecell{45.62 \\ 75.18} & \makecell{58.34 \\ 92.35} & \makecell{56.93 \\ 63.43} \\

\midrule

HuBERT Base~\cite{hsu2021hubert} & \makecell{6.85 \\ 5.41} & \makecell{95.98 \\ 96.30} & \makecell{64.84 \\ 81.42} & \makecell{95.94 \\ 98.34} & \makecell{62.94 \\ 64.92} \\

    \bottomrule
    \end{tabular}}
\label{table:compare_protocols}
\end{table}

We verify that the advantage of our learnable weighted-sum evaluation protocol over the conventional layer-layer frozen representation is consistent across numerous tasks and SSL models, as shown in Table~\ref{table:compare_protocols}.
Note that we do not run this comparison for all the SSL models in Table~\ref{table:upstream_info} due to the computation cost.
Table~\ref{table:compare_protocols} shows that in most cases weighted-sum is better than the last-layer representation, either equally good or significantly better.
Conversely, most of the highlighted failing cases have only slight differences.

\subsection{Full Benchmark Result}
\label{section:full_benchmark_result}

\begin{figure*}[t!]
\centering
\includegraphics[width=6.9in]{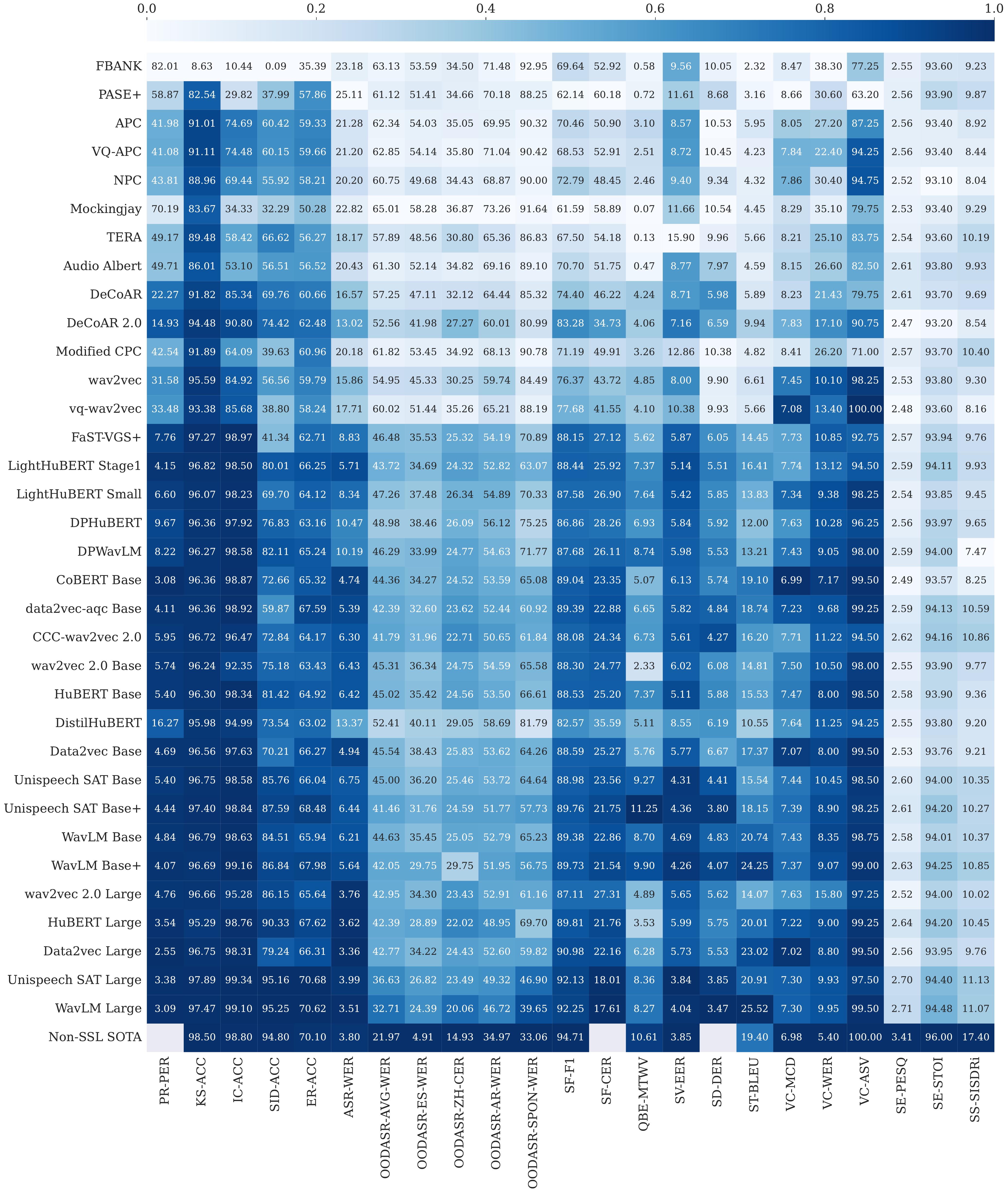}
\caption{
The full benchmark results of 33 foundation models on 15 speech processing tasks.
Each column represents a metric of a task.
The heatmap reflects the performance linearly, and the darker cells of the same task always indicate better performance.
}
\label{fig:benchmark}
\end{figure*}

We present the main results following the weighted-sum protocol in Fig~\ref{fig:benchmark}, and discuss several findings in this section.
The darker color in Fig~\ref{fig:benchmark} suggests the better performance.
Note that our evaluation process involves a fixed downstream model for each task.
\cite{zaiem2023speech} has pointed out using different downstream models may lead to different rankings among speech foundation models.
Hence, our current results only provide the comparison between models in our specific setting.
When the performances between models are too close, we suggest not to rank the models and evaluate them on more corpora and scenarios (e.g. few-shot learning in Section~\ref{section:robustness_of_superb}) to verify the difference.
Conducting statistical tests as illustrated in Section~\ref{section:significance} helps in understanding whether the differences and rankings are meaningful.
A more thorough fix is adopting the \textit{multi-probe} protocol\footnote{\label{multiprob}
The multi-probe protocol combines the results from multiple downstream models, either by choosing the best result or averaging all the results.
}  in ~\cite{zaiem2023speech} or fine-tuning the entire SSL model with only a linear prediction head for each task.
However, both options are computationally infeasible.
We plan to explore these approaches in conjunction with the few-shot learning setting in our future work. Few-shot learning converges faster and reduces training time. Moreover, few-shot benchmarking magnifies performance differences, thereby revealing the true capabilities of foundation models, as demonstrated in Section~\ref{section:robustness_of_superb}.

\subsubsection{SSL outperforms baseline representation (FBANK) on the task generalizability}
\label{section:SSL_vs_FBANK}

We use FBANK\footnote{We use the term “FBANK” to refer to a basic mel-frequency filterbank; in our work we used 80 mel-frequency filters implemented via FFT with a window size of 25 ms and a hop size of 10 ms. We take log energies of the filter outputs.} as the baseline for evaluating the task generalizability due to its wide adoption in most of the speech processing systems.
According to Fig~\ref{fig:benchmark}, all the models outperform FBANK on most tasks.
We parse a few exceptions into the last column of Table~\ref{table:upstream_info} to show the failing cases for each SSL model.
About half of the 33 models outperform FBANK on all tasks.
Another half of the models underperform FBANK on around only 1$\sim$3 tasks out of the full 15 tasks.
The most common failing tasks are SE and SS, involving the SSL models' robustness (to noisy and mixed speech) and the generative capability.

\subsubsection{The leading SSL models show strong task generalizability}
\label{section:leading_SSL_model}

In Figure~\ref{fig:benchmark}, the leading SSL models at the bottom (i.e. wav2vec 2.0, HuBERT, WavLM) outperforms the baseline FBANK (the first row) significantly on all the tasks.
Furthermore, by comparing to conventional leading approaches without the SSL techniques (the last row), we observe that WavLM Large achieve near or even better results across numerous tasks, including KS, IC, SID, ER, ASR, SF, SV, and ST, by just using lightweight downstream models with the frozen SSL encoder.
The results suggest that our weighted-sum protocol is effective and the leading SSL models exhibit strong task generalizability.

\subsubsection{Performance gap on SE and SS}
\label{section:generation_tasks_fail}

In Fig~\ref{fig:benchmark}, we observe that no SSL model is close to the performance of the top systems for SE and SS, namely CMGAN~\cite{abdulatif2022cmgan} and TDANet~\cite{li2022efficient}.
Usually, the competitive systems on these generative tasks use features with stride size smaller than 10 ms.
\cite{huang2022investigating} reported consistent improvements when using smaller strides.
On the other hand, the leading SSL model with distortion and mixture robustness, WavLM Large, uses a large stride size 20 ms, which potentially leads a mismatch in the stride size.
Overall, the results suggest room for future improvement in developing speech foundation models. The current SSL models excel in understanding tasks but lag behind traditional approaches in generative tasks.

\subsubsection{The leading models on VC are different}
\label{section:vc_leading_models}

\begin{table}[t]
    \centering
    \caption{
Comparing VC leading models with their counterparts on the content, speaker, and VC tasks. \textsuperscript{$\ast$}The rows are sorted by MCD. 
}
    \resizebox{0.42\textwidth}{!}{
    \begin{tabular}{@{}lccccc@{}}
        \toprule
        \multirow{2}{*}[-1em]{Models} & PR & SID & \multicolumn{3}{c}{VC} \\
        \cmidrule(lr){2-2} \cmidrule(lr){3-3} \cmidrule(lr){4-6} \\
        {} & per $\downarrow$ & acc $\uparrow$ & mcd\textsuperscript{$\ast$} $\downarrow$ & wer $\downarrow$ & asv $\uparrow$ \\
        \midrule

CoBERT Base~\cite{cobert} & 3.08 & 72.66 & 6.99 & 7.17 & 99.50 \\

Data2vec Base~\cite{data2vec} & 4.69 & 70.21 & 7.07 & 8.00 & 99.50 \\

WavLM Base~\cite{wavlm} & 4.84 & 84.51 & 7.43 & 8.35 & 98.75 \\

HuBERT Base~\cite{hsu2021hubert} & 5.40 & 81.42 & 7.47 & 8.00 & 98.50 \\

\midrule

Data2vec Large~\cite{data2vec} & 2.55 & 79.24 & 7.02 & 8.80 & 99.50 \\

HuBERT Large~\cite{hsu2021hubert} & 3.54 & 90.33 & 7.22 & 9.00 & 99.25 \\

WavLM Large~\cite{wavlm} & 3.09 & 95.25 & 7.30 & 9.95 & 99.50 \\

    \bottomrule
    \end{tabular}}
\label{table:compare_vc_leading_models}
\end{table}

Surprisingly, WavLM series do not reach the first rank on the VC task.
Instead, the leading models are CoBERT Base~\cite{cobert} and Data2vec Base/Large~\cite{data2vec}, which show competitive results on all 3 metrics of VC (MCD, WER, ASV-acc).
We compare these models with their similar-sized while slightly worse competitors HuBERT and WavLM in Table~\ref{table:compare_vc_leading_models}.
We use MCD as the primary sorting metric since it demonstrates much higher correlation with human perception in terms of naturalness and speaker similarity in \cite{huang2022s3prl}.
Table~\ref{table:compare_vc_leading_models} shows that the VC leading models learn better speaker-independent content representation\textsuperscript{\ref{footnote:speaker_invariance}}, aligning the results in the VC field~\cite{contentvec,qian2019autovc,hsu2016voice,hsu2017voice}.
Specifically, they possess higher content accessibility\footnote{\label{footnote:accessibility} We define \textit{accessibility} as \textit{how easy we can extract the specific type of information by a shallow classifier}. The definition follows the conventional works in speech SSL~\cite{apc1,vq_apc,mockingjay}} and speaker invariance\textsuperscript{\ref{footnote:speaker_invariance}}, as indicated by the better PER and poorer speaker accuracy, which help generalize to unseen \textit{source} speakers during the conversion stage.
Intuitively, it seems impossible to perform well on the content, speaker and the VC task simultaneously with the same foundation model.
However, we will show in Section~\ref{section:single_layer_for_vc} that it is feasible.

\section{Layer-wise Analysis on SUPERB}
\label{section:layer-wise-analysis}

In this section, we hypothesize that distinct layers are responsible for different tasks and examine their individual contributions.

\subsection{Layer-weights are not proportional to layer performances}
\label{section:layer-weights-analysis}

\begin{figure}[t!]
\centering
\includegraphics[width=3.5in]{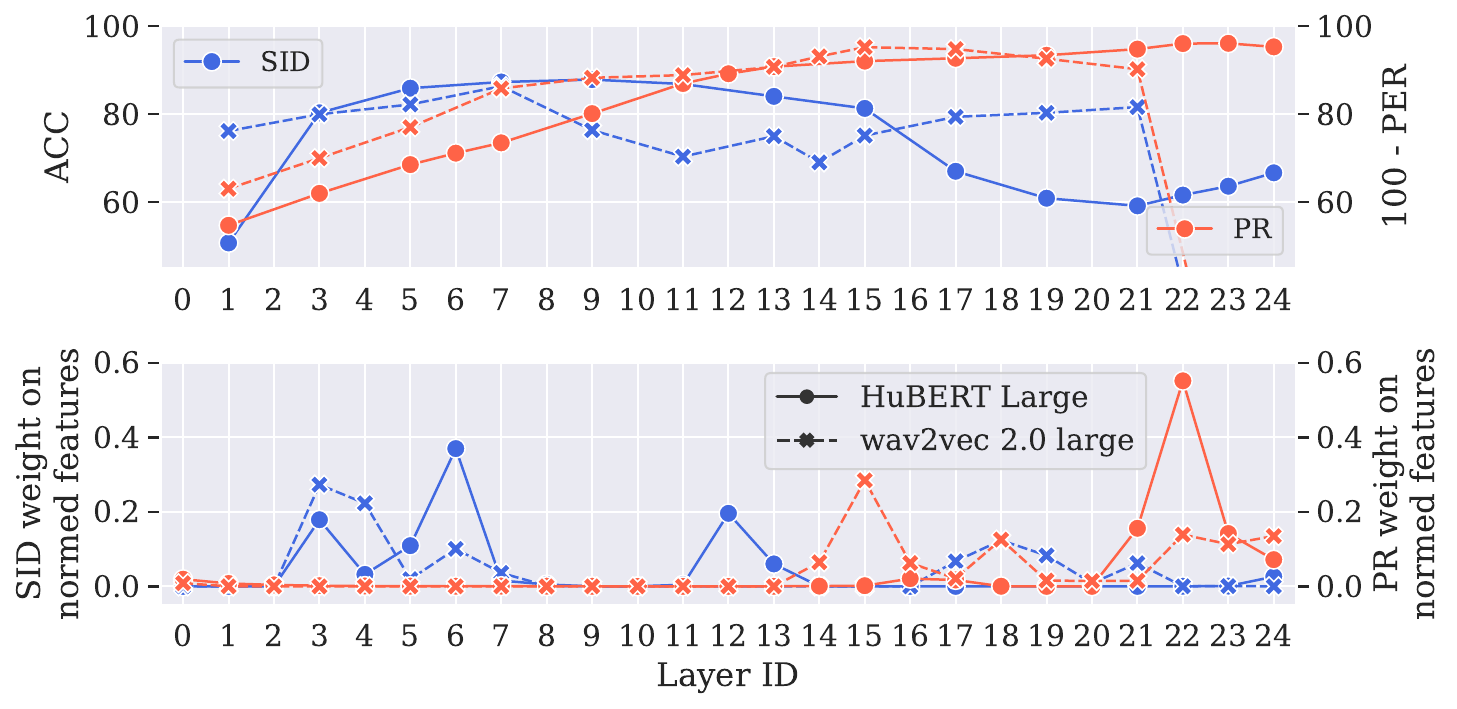}
\caption{
Comparing each layer's performance to layer-weights after the weighted-sum benchmarking. The blue lines are for SID; the red lines are for PR. The solid lines are for HuBERT Large; the dashed lines are for wav2vec 2.0 Large.
}
\label{fig:check_layer_weights}
\end{figure}

It is widely adopted to analyze each layer's importance to a task by layer weights\textsuperscript{\ref{footnote:layer_weight}}~\cite{wavlm,chang2022distilhubert,shi2023ml}, under the hypothesis that the weights are proportional to each layer's true performance. However, our analyses suggest that layer weights are not informative.

\begin{figure}[t!]
\centering
\includegraphics[width=3.2in]{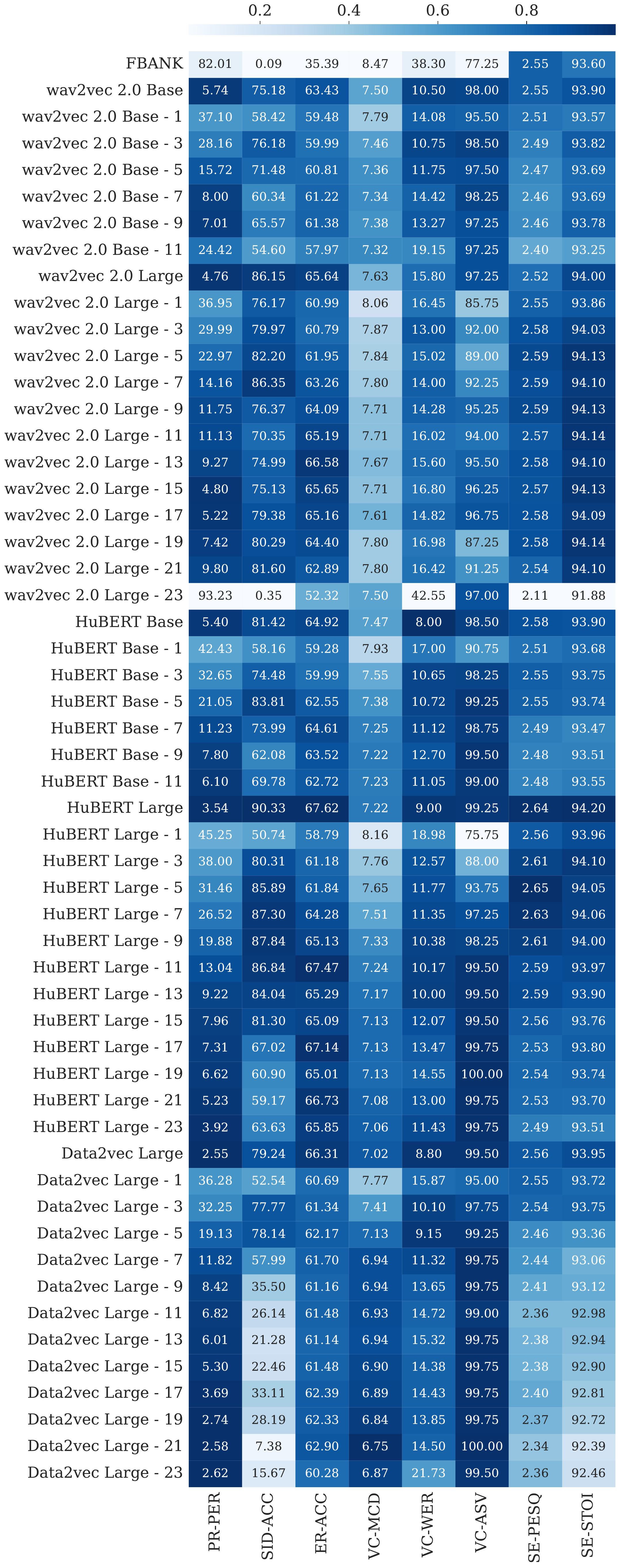}
\caption{
Layer-wise benchmarking of wav2vec 2.0 Large, HuBERT Large and Data2vec Large on 5 tasks.
The model name's suffix number represents the layer ID.
The row without the layer ID is the weighted-sum result.
}
\label{layer-wise-benchmark}
\end{figure}

Some models possess different numerical scales across layers, which essentially affect the layer weights and the interpretation in our preliminary experiments
\footnote{
This phenomenon is especially observable for the Large variants of wav2vec 2.0, HuBERT and WavLM.
These models' last layer features are in very small numerical values compared to the other layers, and the layer-weight of the last layer is extremely large.
However, the last layer of wav2vec 2.0 does not contain useful information according to Fig~\ref{fig:check_layer_weights} and Fig~\ref{layer-wise-benchmark}.
Hence, we infer that the layer-weights might serve two functionalities jointly: (1) normalizing the numerical scale across layers, (2) identifying the informative layers.
As an importance analysis tool, we only care about the functionality (2), hence the raw layer-weight from the default benchmarking is not an appropriate choice.
}.
Therefore, we consider another benchmarking setting to factor out the effect of the feature numerical scale: \textit{normalized benchmarking}. In normalized benchmarking, we first normalize each layer of features by a layer norm across the hidden size dimension and then take the \textit{normalized features} for the benchmarking.

The results for wav2vec 2.0 Large and HuBERT Large are presented in Fig~\ref{fig:check_layer_weights}.\footnote{Switching to the normalized benchmarking does not affect the task performance, but is more rigorous for analyzing the layer-weights.}
We show layer-weights for all the layers, while due to the huge computation cost of layer-wise benchmarking, we only benchmark the odd layers.
The layer-weights can only roughly reflect the true performance on PR and SID with many inconsistencies.
For example, the layer-weights on SID fail to locate the best layer for both wav2vec 2.0 Large and HuBERT Large.
Furthermore, on both PR and SID the layer-weights fail to reflect the smooth information change inside the speech foundation models.

\begin{table}[t]
    \centering
    \caption{Spearman's $\rho$ between layer-weights from a \textit{normalized benchmarking} and the true layer performances. The \textit{Score} is designed to be higher for better performance.}
    \begin{tabular}{cccccc}
        \toprule
        Task & PR & SID & ER & VC & SE \\
        \cmidrule(lr){1-1} \cmidrule(lr){2-2} \cmidrule(lr){3-3} \cmidrule(lr){4-4} \cmidrule(lr){5-5} \cmidrule(lr){6-6} \\
        Score & 100 - per & acc & acc & -mcd & pesq \\
        \midrule
        $\rho$ & 0.393 & 0.494 & 0.371 & -0.693 & 0.711 \\
        p-value & 0.031 & 0.007 & 0.041 & 0 & 0 \\
        \bottomrule
    \end{tabular}
    \label{table:layer_weights_different_tasks}
\end{table}

Quantitatively, we compute the Spearman’s rank correlation coefficient (Spearman's $\rho$)~\cite{pasad2021layer} between the layer performances and the layer-weights.
PR, SID, ER, VC and SE are examined.
Table~\ref{table:layer_weights_different_tasks} shows that the layer-weights are not proportional to layer performances for all the tasks except SE.
As a result, we suggest to conduct layer-wise single-layer benchmarking for assessing each layer's quality for each task.

\subsection{Layer-wise single-layer benchmarking}
\label{layerwise_benchmarking}

In Fig~\ref{layer-wise-benchmark}, we present the single-layer benchmark results for some representative models: wav2vec 2.0 Base/Large, HuBERT Base/Large, and Data2vec Large.
Due to the huge computational cost, we benchmark the odd layers on a subset of representative tasks: PR, SID, ER, VC and SE.

\subsubsection{Different tasks favor different layers}
\label{section:information_flow}

The accessibility\textsuperscript{\ref{footnote:accessibility}} of information in the internal layers shows a similar trend across different models. The lower layers benefit the SE task, which requires manipulating STFT masks\footnote{The results are evident when considering HuBERT Base/Large and Data2vec Large. For wav2vec 2.0 Base/Large, the SE differences between layers are less obvious.}, while the middle layers are more beneficial for the speaker task SID, followed by the prosody-centric ER. Finally, the higher layers are more beneficial for the content task PR. As a result, no single-layer representation can achieve competitive performance on all tasks, but it is feasible to achieve the goal by training a learnable weighted-sum to ensemble all the layers.

\subsubsection{Pushing limits with single-layer benchmarking}
\label{section:pushing_limits}

We observe that the best single-layer benchmarking result sometimes outperforms the weighted-sum benchmarking result, with wav2vec 2.0 Base and HuBERT Base on SID, wav2vec 2.0 Large on ER, wav2vec 2.0 Large and HuBERT Large for SE, and all the models on VC with the MCD metric as examples. This phenomenon has also been reported in~\cite{pasad2023comparative}, which analyzes the content information across layers. Our results suggest that sweeping the internal layers can potentially lead to improved performances, and this phenomenon is consistent across several tasks. However, since the improvement observed in the single-layer benchmarking is more related to individual SSL models and is not consistent across all models except VC, we still use the weighted-sum approach as the default benchmarking protocol. Additionally, the weighted-sum protocol requires much less computation cost since it only requires a single run.

\subsubsection{Speaker-independent representation}
\label{section:speaker_independent}

We observe that some SSL models possess the speaker-independent representation at their final layers.
Following the analyses in Section~\ref{section:vc_leading_models}, when considering the relation between the content task PR, the speaker task SID and VC across layers, we find that HuBERT Base/Large and Data2vec Large show a higher speaker invariance at their final layers.
These layers possess higher content accessibility and lower speaker accessibility.
Compared to HuBERT Large, wav2vec 2.0 Large shows poorer speaker invariance at all layers, which is evident in Fig~\ref{fig:check_layer_weights}.
Consequently, the best individual layer for VC in HuBERT Large outperforms that in wav2vec 2.0 Large, with 7.06 MCD and 7.5 MCD.
Furthermore, when comparing to Data2vec Large, we observe Data2vec Large achieves the highest degree of speaker invariance (7.38 ACC) with the best phonetic information (2.58 PER), and reach the best VC performance with 6.75 MCD and 100.00 ACC on target speaker similarity\footnote{
We also conduct the layer-wise benchmarking for WavLM Large, and the results are similar to those of HuBERT Large. The best layer for VC achieve 7.06 MCD, 10.85 WER, and 98.25 target speaker accuracy, which is also worse than Data2vec Large. We do not show all the results due to space limit.
}.

\subsubsection{Single-layer benchmarking for VC}
\label{section:single_layer_for_vc}

The results in Section~\ref{section:speaker_independent} suggest that one should consider the single-layer benchmarking on VC, since the recognize-synthesis framework essentially prefers the representation independent of the speaker variations\textsuperscript{\ref{footnote:speaker_invariance}}.
When all the layers are used, the layers of high speaker accessibility inevitably leads to the source speaker leakage and misguide the VC training.
This modification enables the possibility for a speech foundation to excel on all the SUPERB tasks, as long as it possesses layers with high speaker accessibility and a separate set of layers with high content accessibility and speaker invariance.
As a result, we suggest conducting the weighted-sum approach for most of the SUPERB tasks except QbE and VC.
The former relies on the non-trainable DTW algorithm; the latter needs to exclude the source speaker information for technical correctness.

\section{Statistical significance in SUPERB}
\label{section:significance}

\begin{table}[ht]
    \centering
    \caption{
The p-values for comparing the Large models.
The bold cells mark the cases when the difference is insignificant~(p-value \textgreater\ 0.05).
}
    \resizebox{0.41\textwidth}{!}{
    \begin{tabular}{ccccccc}
        \toprule
        {} & {} & HuBERT & W2V2 & Data2vec & WavLM & {} \\
        \midrule
        {} & per & \multicolumn{2}{c}{\cellcolor{blue!15}{PR}} & \multicolumn{2}{c}{\cellcolor{orange!25}{ASR}} & wer \\
        \cmidrule(lr){2-2} \cmidrule(lr){3-6} \cmidrule(lr){7-7} \\
        HuBERT & 3.29 & \cellcolor{gray!80}{\texttimes} & \cellcolor{orange!25}{\textbf{.0610}} & \cellcolor{orange!25}{0} & \cellcolor{orange!25}{0} & 3.76 \\
        W2V2 & 4.75 & \cellcolor{blue!15}{0} & \cellcolor{gray!80}{\texttimes} & \cellcolor{orange!25}{0} & \cellcolor{orange!25}{0} & 3.62 \\
        Data2vec & 2.55 & \cellcolor{blue!15}{0} & \cellcolor{blue!15}{0} & \cellcolor{gray!80}{\texttimes} & \cellcolor{orange!25}{\textbf{.2260}} & 3.44 \\
        WavLM & 3.22 & \cellcolor{blue!15}{\textbf{.0930}} & \cellcolor{blue!15}{0} & \cellcolor{blue!15}{0} & \cellcolor{gray!80}{\texttimes} & 3.36 \\
        \midrule
        {} & acc & \multicolumn{2}{c}{\cellcolor{red!15}{KS}} & \multicolumn{2}{c}{\cellcolor{green!20}{QbE}} & mtwv \\ \\
        \cmidrule(lr){2-2} \cmidrule(lr){3-6} \cmidrule(lr){7-7} \\
        HuBERT & 95.29 & \cellcolor{gray!80}{\texttimes} & \cellcolor{green!20}{\textbf{.1192}} & \cellcolor{green!20}{.0018} & \cellcolor{green!20}{0} & 3.53 \\
        W2V2 & 96.27 & \cellcolor{red!15}{.009} & \cellcolor{gray!80}{\texttimes} & \cellcolor{green!20}{.0174} & \cellcolor{green!20}{0} & 5.06 \\
        Data2vec & 96.75 & \cellcolor{red!15}{0} & \cellcolor{red!15}{\textbf{.1289}} & \cellcolor{gray!80}{\texttimes} & \cellcolor{green!20}{0} & 6.28 \\
        WavLM & 97.47 & \cellcolor{red!15}{0} & \cellcolor{red!15}{0} & \cellcolor{red!15}{0} & \cellcolor{gray!80}{\texttimes} & 8.86 \\
        \midrule
        {} & acc & \multicolumn{2}{c}{\cellcolor{blue!15}{IC}} & \multicolumn{2}{c}{\cellcolor{orange!25}{ER}} & acc \\
        \cmidrule(lr){2-2} \cmidrule(lr){3-6} \cmidrule(lr){7-7} \\
        HuBERT & 98.76 & \cellcolor{gray!80}{\texttimes} & \cellcolor{orange!25}{.0028} & \cellcolor{orange!25}{.0005} & \cellcolor{orange!25}{.0354} & 67.58 \\
        W2V2 & 95.68 & \cellcolor{blue!15}{0} & \cellcolor{gray!80}{\texttimes} & \cellcolor{orange!25}{\textbf{.5558}} & \cellcolor{orange!25}{0} & 65.64 \\
        Data2vec & 98.31 & \cellcolor{blue!15}{\textbf{.0827}} & \cellcolor{blue!15}{0} & \cellcolor{gray!80}{\texttimes} & \cellcolor{orange!25}{0} & 65.29 \\
        WavLM & 99.31 & \cellcolor{blue!15}{.0035} & \cellcolor{blue!15}{0} & \cellcolor{blue!15}{0} & \cellcolor{gray!80}{\texttimes} & 68.87 \\
        \midrule
        {} & slot-f1 & \multicolumn{2}{c}{\cellcolor{red!15}{SF}} & \multicolumn{2}{c}{\cellcolor{green!20}{SD}} & der \\
        \cmidrule(lr){2-2} \cmidrule(lr){3-6} \cmidrule(lr){7-7} \\
        HuBERT & 89.81 & \cellcolor{gray!80}{\texttimes} & \cellcolor{green!20}{\textbf{.1569}} & \cellcolor{green!20}{\textbf{.0513}} & \cellcolor{green!20}{0} & 5.75 \\
        W2V2 & 86.94 & \cellcolor{red!15}{0} & \cellcolor{gray!80}{\texttimes} & \cellcolor{green!20}{\textbf{.5412}} & \cellcolor{green!20}{0} & 5.62 \\
        Data2vec & 90.98 & \cellcolor{red!15}{0} & \cellcolor{red!15}{0} & \cellcolor{gray!80}{\texttimes} & \cellcolor{green!20}{0} & 5.53 \\
        WavLM & 92.21 & \cellcolor{red!15}{0} & \cellcolor{red!15}{0} & \cellcolor{red!15}{.0001} & \cellcolor{gray!80}{\texttimes} & 3.24 \\
        \midrule
        {} & acc & \multicolumn{2}{c}{\cellcolor{blue!15}{SID}} & \multicolumn{2}{c}{\cellcolor{orange!25}{SV}} & eer \\
        \cmidrule(lr){2-2} \cmidrule(lr){3-6} \cmidrule(lr){7-7} \\
        HuBERT & 90.33 & \cellcolor{gray!80}{\texttimes} & \cellcolor{orange!25}{.0386} & \cellcolor{orange!25}{\textbf{.1212}} & \cellcolor{orange!25}{0} & 5.99 \\
        W2V2 & 86.15 & \cellcolor{blue!15}{0} & \cellcolor{gray!80}{\texttimes} & \cellcolor{orange!25}{\textbf{.6039}} & \cellcolor{orange!25}{0} & 5.65 \\
        Data2vec & 76.77 & \cellcolor{blue!15}{0} & \cellcolor{blue!15}{0} & \cellcolor{gray!80}{\texttimes} & \cellcolor{orange!25}{0} & 5.73 \\
        WavLM & 95.49 & \cellcolor{blue!15}{0} & \cellcolor{blue!15}{0} & \cellcolor{blue!15}{.0002} & \cellcolor{gray!80}{\texttimes} & 3.77 \\
        \midrule
        {} & wer & \multicolumn{2}{c}{\cellcolor{red!15}{OOD-ASR~(avg)}} & \multicolumn{2}{c}{\cellcolor{green!20}{ST}} & bleu \\
        \cmidrule(lr){2-2} \cmidrule(lr){3-6} \cmidrule(lr){7-7} \\
        HuBERT & 42.28 & \cellcolor{gray!80}{\texttimes} & \cellcolor{green!20}{0} & \cellcolor{green!20}{0} & \cellcolor{green!20}{0} & 20.23 \\
        W2V2 & 42.90 & \cellcolor{red!15}{0} & \cellcolor{gray!80}{\texttimes} & \cellcolor{green!20}{0} & \cellcolor{green!20}{0} & 12.78 \\
        Data2vec & 42.71 & \cellcolor{red!15}{0} & \cellcolor{red!15}{\textbf{.3103}} & \cellcolor{gray!80}{\texttimes} & \cellcolor{green!20}{.0113} & 23.02 \\
        WavLM & 32.66 & \cellcolor{red!15}{0} & \cellcolor{red!15}{0} & \cellcolor{red!15}{0} & \cellcolor{gray!80}{\texttimes} & 26.56 \\
        \midrule
        {} & pesq & \multicolumn{2}{c}{\cellcolor{blue!15}{SE~(pesq)}} & \multicolumn{2}{c}{\cellcolor{orange!25}{SE~(stoi)}} & stoi \\
        \cmidrule(lr){2-2} \cmidrule(lr){3-6} \cmidrule(lr){7-7} \\
        HuBERT & 94.18 & \cellcolor{gray!80}{\texttimes} & \cellcolor{orange!25}{0} & \cellcolor{orange!25}{0} & \cellcolor{orange!25}{0} & 2.64 \\
        W2V2 & 94.04 & \cellcolor{blue!15}{.0036} & \cellcolor{gray!80}{\texttimes} & \cellcolor{orange!25}{0} & \cellcolor{orange!25}{0} & 2.52 \\
        Data2vec & 93.95 & \cellcolor{blue!15}{0} & \cellcolor{blue!15}{.0444} & \cellcolor{gray!80}{\texttimes} & \cellcolor{orange!25}{0} & 2.56 \\
        WavLM & 94.51 & \cellcolor{blue!15}{0} & \cellcolor{blue!15}{0} & \cellcolor{blue!15}{.0002} & \cellcolor{gray!80}{\texttimes} & 2.70 \\
        \midrule
        {} & sisdri & \multicolumn{2}{c}{\cellcolor{red!15}{SS}} & \multicolumn{2}{c}{\cellcolor{green!20}{VC}} & mcd \\
        \cmidrule(lr){2-2} \cmidrule(lr){3-6} \cmidrule(lr){7-7} \\
        HuBERT & 10.45 & \cellcolor{gray!80}{\texttimes} & \cellcolor{green!20}{0} & \cellcolor{green!20}{0} & \cellcolor{green!20}{0} & 7.22 \\
        W2V2 & 10.02 & \cellcolor{red!15}{0} & \cellcolor{gray!80}{\texttimes} & \cellcolor{green!20}{0} & \cellcolor{green!20}{0} & 7.63 \\
        Data2vec & 9.76 & \cellcolor{red!15}{0} & \cellcolor{red!15}{0} & \cellcolor{gray!80}{\texttimes} & \cellcolor{green!20}{0} & 7.02 \\
        WavLM & 11.07 & \cellcolor{red!15}{0} & \cellcolor{red!15}{0} & \cellcolor{red!15}{0} & \cellcolor{gray!80}{\texttimes} & 7.3 \\
        \midrule
        {} & wer & \multicolumn{2}{c}{\cellcolor{blue!15}{OOD-ASR~(es)}} & \multicolumn{2}{c}{\cellcolor{orange!25}{OOD-ASR~(ar)}} & wer \\
        \cmidrule(lr){2-2} \cmidrule(lr){3-6} \cmidrule(lr){7-7} \\
        HuBERT & 28.89 & \cellcolor{gray!80}{\texttimes} & \cellcolor{orange!25}{0} & \cellcolor{orange!25}{0} & \cellcolor{orange!25}{0} & 48.95 \\
        W2V2 & 34.3 & \cellcolor{blue!15}{0} & \cellcolor{gray!80}{\texttimes} & \cellcolor{orange!25}{\textbf{.3680}} & \cellcolor{orange!25}{0} & 52.91 \\
        Data2vec & 34.22 & \cellcolor{blue!15}{0} & \cellcolor{blue!15}{\textbf{.8730}} & \cellcolor{gray!80}{\texttimes} & \cellcolor{orange!25}{0} & 52.6 \\
        WavLM & 24.39 & \cellcolor{blue!15}{0} & \cellcolor{blue!15}{0} & \cellcolor{blue!15}{0} & \cellcolor{gray!80}{\texttimes} & 46.72 \\
        \midrule
        {} & cer & \multicolumn{2}{c}{\cellcolor{red!15}{OOD-ASR~(zh)}} & \multicolumn{2}{c}{\cellcolor{green!20}{OOD-ASR~(spon)}} & wer \\
        \cmidrule(lr){2-2} \cmidrule(lr){3-6} \cmidrule(lr){7-7} \\
        HuBERT & 22.02 & \cellcolor{gray!80}{\texttimes} & \cellcolor{green!20}{0} & \cellcolor{green!20}{0} & \cellcolor{green!20}{0} & 69.7 \\
        W2V2 & 23.43 & \cellcolor{red!15}{0} & \cellcolor{gray!80}{\texttimes} & \cellcolor{green!20}{0} & \cellcolor{green!20}{0} & 61.16 \\
        Data2vec & 24.43 & \cellcolor{red!15}{0} & \cellcolor{red!15}{0} & \cellcolor{gray!80}{\texttimes} & \cellcolor{green!20}{0} & 59.82 \\
        WavLM & 20.06 & \cellcolor{red!15}{0} & \cellcolor{red!15}{0} & \cellcolor{red!15}{0} & \cellcolor{gray!80}{\texttimes} & 39.65 \\
        \bottomrule
    \end{tabular}}
    \label{table:significance}
\end{table}

We analyze the statistical significance of the SUPERB leaderboard since it contains highly similar results among the top-performing models while the significance of the improvement remains unknown.
To compute the p-values, we use the \textit{sclite}\footnote{\href{https://github.com/usnistgov/SCTK}{https://github.com/usnistgov/SCTK}} toolkit to conduct the MAPSSWE~\cite{pallet1990tools} test for PR, ASR, and the OOD-ASR tasks.
For SV, we conduct the proportional test on the corresponding classification error following \cite{bengio04b_odyssey}.
For ST, we follow \cite{koehn2004statistical} to conduct paired bootstrap resampling.
For the classification tasks KS, IC, ER, SID, we conduct the McNemar test~\cite{kornblith2019better}.
We conduct the Student's t-test for SD~\cite{lin2019lstm}.
For the remaining tasks VC, SE, SS, and QbE, since no apparent choice was found to our best knowledge, we conduct the Student's t-test.

We present the results in Table~\ref{table:significance} for four leading SSL models: wav2vec 2.0 Large, HuBERT Large, Data2vec Large and WavLM Large.
On most of the tasks the model differences are significant, while the differences in SV and SD are frequently insignificant.
In terms of DER scores, Data2vec Large ranks ahead of wav2vec 2.0 Large, followed by HuBERT Large.
However, the p-values indicate that their performances are statistically equal.
The results suggest that statistical significance should be considered when ranking models, as even a minor random disturbance can result in a noticeable alteration in the ranking.
On QBE, wav2vec 2.0 Large and Data2vec Large show a 1.22 MTWV difference which is significant, while wav2vec 2.0 Large and HuBERT Large show a 1.53 MTWV difference which is instead insignificant, suggesting that a larger difference on the overall scores do not necessarily lead to more significant results.
Despite that the PESQ, STOI and SISDRi scores on SE and SS are highly similar for all the models, they all pass the significance test, suggesting that the improvement is small but consistent across the testing utterances.
We conclude that statistically insignificant results exist and encourage the participants to conduct statistical tests.
We will release the downstream prediction files, along with the recipes for calculating the p-values.

\section{Robustness of SUPERB}
\label{section:robustness_of_superb}

We discuss the robustness of the proposed benchmark.
We examined the robustness of SUPERB-SG in ~\cite{tsai2022superb}, and extend the examination to the tasks defined in SUPERB~\cite{superb} in this work.
Due to the space limit, we select representative tasks, PR, SID and ER for content, speaker and paralinguistic information respectively.

\begin{figure}[t!]
\centering

\begin{subfigure}{0.22\textwidth}
    \includegraphics[width=\textwidth]{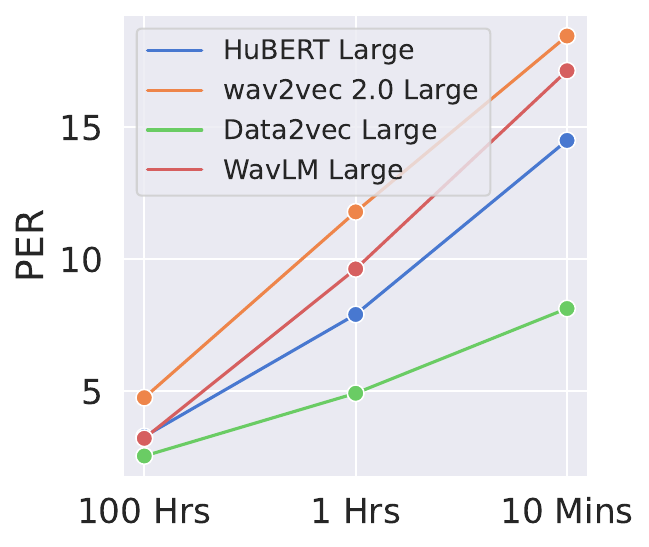}
    \caption{Few-shot PR}
    \label{fig:few_PR}
\end{subfigure}
\hfill
\begin{subfigure}{0.22\textwidth}
    \includegraphics[width=\textwidth]{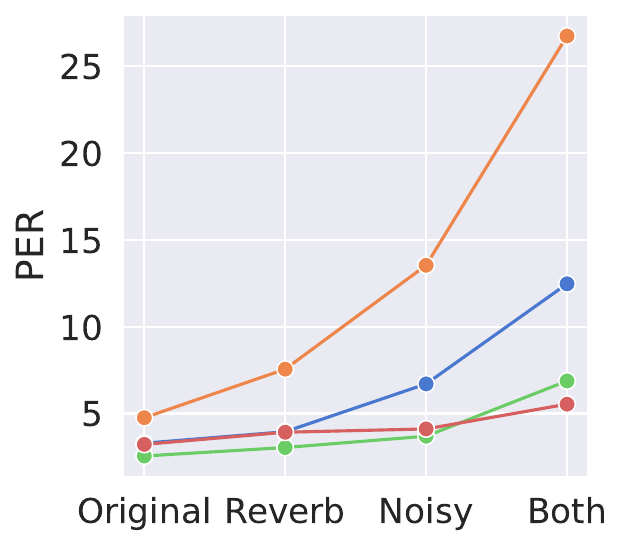}
    \caption{Distorted PR}
    \label{fig:distort_PR}
\end{subfigure}

\begin{subfigure}{0.22\textwidth}
    \includegraphics[width=\textwidth]{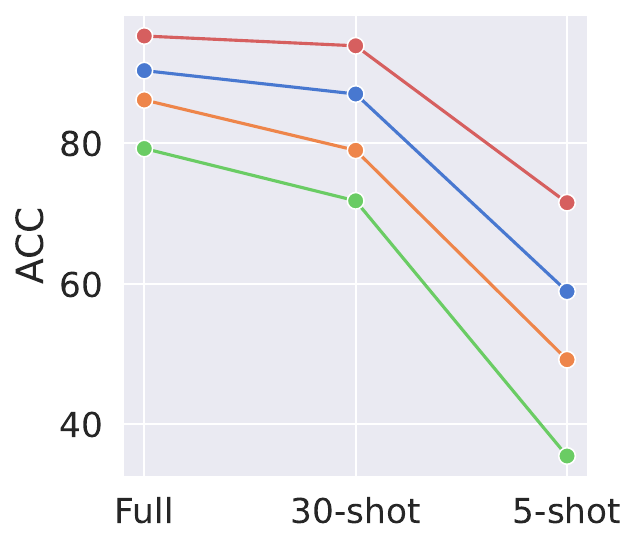}
    \caption{Few-shot SID}
    \label{fig:few_SID}
\end{subfigure}
\hfill
\begin{subfigure}{0.22\textwidth}
    \includegraphics[width=\textwidth]{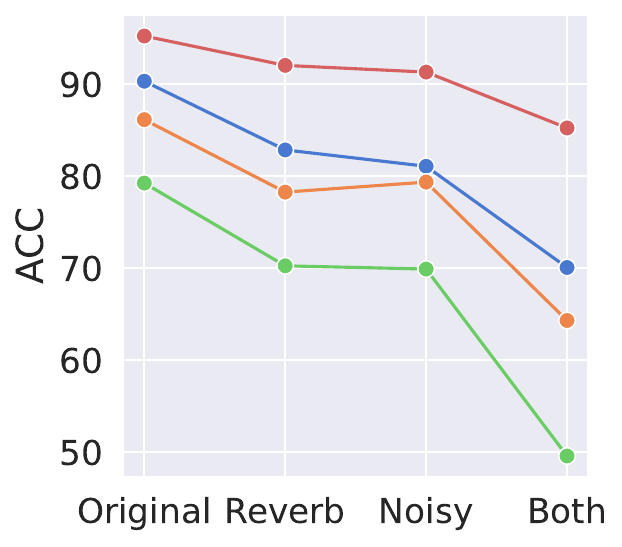}
    \caption{Distorted SID}
    \label{fig:distort_SID}
\end{subfigure}

\begin{subfigure}{0.22\textwidth}
    \includegraphics[width=\textwidth]{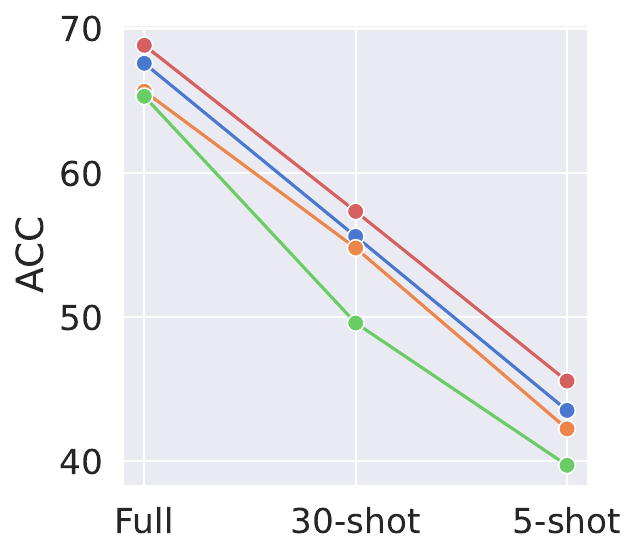}
    \caption{Few-shot ER}
    \label{fig:few_ER}
\end{subfigure}
\hfill
\begin{subfigure}{0.22\textwidth}
    \includegraphics[width=\textwidth]{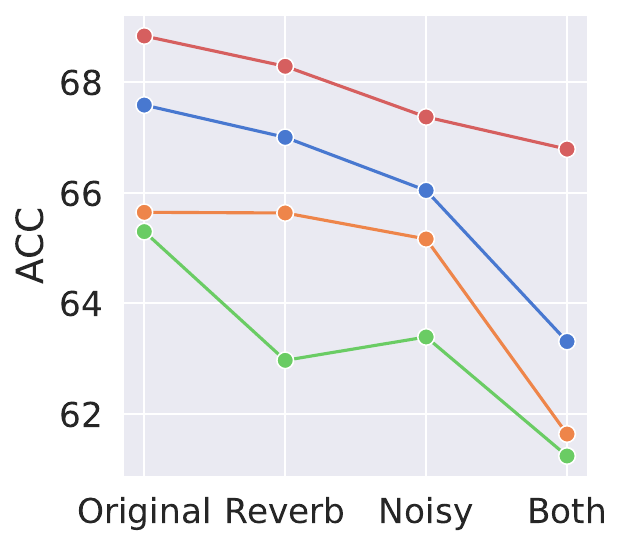}
    \caption{Distorted ER}
    \label{fig:distort_ER}
\end{subfigure}
\caption{
PR, SID and ER under few-shot and distorted conditions with HuBERT Large, wav2vec 2.0 Large, Data2vec Large and WavLM Large.
}
\label{robustness}
\end{figure}

We discuss two types of condition variations: low-resource and distorted recordings.
For the low-resource condition, we consider two levels.
For PR, we randomly sample 1 hour and 10 minutes of recordings from the LibriSpeech train-clean-100 subset for the few-shot and extreme few-shot conditions respectively; we randomly sample 30 and 5 utterances from each speaker for SID; we randomly sample 30 and 5 utterances from each emotion category for ER.
The development and the testing sets are the same as the original ones.
For the distorted condition, we consider applying additive noises or/and reverberation to the training, development and testing sets.
For additive noise, the WHAM!~\cite{wichern2019wham} dataset's training, validation, and testing sets are applied to the training, validation, and testing sets of PR, SID and ER respectively.
The SNR for each noise addition is randomly sampled from 3, 6, and 9 dB.
For reverberation, we convolve the speech with the room impulse response (RIR) data in \cite{ko2017study}.
The simulated RIR are divided into training and development sets for the corresponding speech.
The real RIR is applied to the testing speech.
When both additive noises and reverberation are applied, we follow the same settings above, with additive noises followed by reverberation.
We present the results for HuBERT Large, wav2vec 2.0 Large, WavLM Large and Data2vec Large in Fig~\ref{robustness}.

Firstly, Fig~\ref{robustness} shows that for PR, SID and ER, different condition changes do not lead to significantly different rankings.
In the few-shot PR, we find that Data2vec Large shows better robustness in the low-resource conditions according to its smoother slope compared to all the others.
Despite WavLM Large and HuBERT Large showing insignificant difference (Table~\ref{table:significance}) in the default PR setting with 100 hours of data, HuBERT Large is more robust against the few-shot 1-hour and 10-minute settings.
On the other hand, WavLM is more robust against the distorted conditions, and further surpasses Data2vec when both noise and reverberation are applied.
This result suggests that while the models might achieve similar scores in the default SUPERB setting due to the saturating performances, they could possess different robustness characteristics.
In SID and ER, the default SUPERB can perfectly reflect the performance.

In conclusion, similar to the results in \cite{tsai2022superb}, the default experimental settings of SUPERB are robust against various scenarios, albeit with a few exceptions.
Our analysis reveals that each model displays varying degrees of resilience under different conditions.
The standard SUPERB evaluation might not fully capture these nuances.
This finding guides us towards developing a more challenging version of SUPERB.

\section{Conclusion}

We present SUPERB benchmark, a framework for evaluating speech foundation models.
The standardized 15 tasks cover a wide range of speech processing tasks, including both discriminative and generative tasks.
The 33 evaluated models provide comprehensive baselines.
With our extensive evaluations, we verify that SSL models are universal across numerous SUPERB tasks, and the best performing model achieve near or better performances compared to conventional pipelines.
For the benchmark best practice, we suggest to conduct layer-wise single-layer benchmarking for voice conversion due to the speaker invariance property\textsuperscript{\ref{footnote:speaker_invariance}}.
In addition, we observe that layer-weights are not suitable for analyzing layer performances, and the ranking between models requires careful statistical tests.
Finally, our robustness analysis suggests that the distorted and few-shot conditions help create a more challenging and realistic benchmark for general speech understanding and generation.
We open-source all the materials to lower the barrier for reproduction, benchmarking, submission, and analysis.
We welcome researchers to join our active community and drive the research frontier together.

\section*{Acknowledgments}

I would like to express my deepest gratitude to Jardin Hsu, who has always been my most solid spiritual support and guide. Without her, I could not have overcome all the technical and mental challenges to finalize this article. I would also like to express my deepest gratitude to my classical piano teacher, Yiin-bin Yang. Her dedication to music and relentless pursuit of better sound and meaningful messages profoundly influenced me to rethink the purpose and methods of scientific research. Finally, I would like to thank Prof. Lin-shan Lee, whose talks and spirit always deeply inspire me. Thanks National Center for High-performance Computing (NCHC) of National Applied Research Laboratories
(NARLabs) in Taiwan for providing computational and storage
resources.

\bibliographystyle{IEEEtran} 
\bibliography{superb}

\end{document}